\documentclass[aps,pre,reprint,groupedaddress,longbibliography]{revtex4-1}

\usepackage[dvipdfmx]{graphicx}% Include figure files
\usepackage{dcolumn}% Align table columns on decimal point
\usepackage{bm}% bold math
\usepackage{accents}
\usepackage{amssymb}
\usepackage{feynmp}
\usepackage{url}
\usepackage{setspace}
\usepackage{color}
\usepackage{pifont}
\usepackage{hyperref}

\newcommand{\revision}[1]{{{#1}}}

\newcommand{\beq}{\begin{equation}}
\newcommand{\eeq}{\end{equation}}
\newcommand{\beqa}{\begin{eqnarray}}
\newcommand{\eeqa}{\end{eqnarray}}
\newcommand{\Tr}{\text{Tr}}

\begin{document}

%\preprint{}

\title{Symmetry induced enhancement in finite-time thermodynamic trade-off relations}

\author{Ken Funo}
\affiliation{Department of Applied Physics, The University of Tokyo, 7-3-1 Hongo, Bunkyo-ku, Tokyo 113-8656, Japan}
\email{funo@ap.t.u-tokyo.ac.jp}

\author{Hiroyasu Tajima}
\affiliation{Graduate School of Informatics and Engineering, The University of Electro-Communications, 1-5-1 Chofugaoka, Chofu, Tokyo 182-8585, Japan}





\date{\today}

\begin{abstract}
Symmetry imposes constraints on open quantum systems, affecting the dissipative properties in nonequilibrium processes. Superradiance is a typical example in which the decay rate of the system is enhanced via a collective system-bath coupling that respects permutation symmetry. Such model has also been applied to heat engines. However, a generic framework that addresses the impact of symmetry in finite-time thermodynamics is not well established. Here, we show a symmetry-based framework that describes the fundamental limit of collective enhancement in finite-time thermodynamics. Specifically, we derive a general upper bound on the average jump rate, which quantifies the fundamental speed set by thermodynamic speed limits and trade-off relations. We identify the symmetry condition which achieves the obtained bound, and explicitly construct an open quantum system model that goes beyond the enhancement realized by the conventional superradiance model. 
\end{abstract}

\maketitle

\textit{Introduction.---} 
Symmetry plays a fundamental role in physics, as it provides a powerful tool to analyze, classify, and design the system of interest, and imposes constraints on physical systems such as superselection rules, charge conservations, and so forth. 
Realistic systems are unavoidably open to their surrounding degrees of freedom, and hence recent studies~\cite{Buca12, SymmetryConservedQuantity, nonHermitian, nonHermitian1, Bardyn_2013, Ryu23} aim to understand the impact of symmetry in open quantum systems. Consequently, discussions based on the symmetry of the Hamiltonian have been extended to those based on non-Hermitian Hamiltonians~\cite{nonHermitian_review, Bender_2007} and Gorini-Kossakowski-Sudarshan-Lindblad (GKSL) master equations~\cite{Lindblad, GKS, Breuer}.  
Symmetry is also related to topology~\cite{nonHermitian, nonHermitian1, Bardyn_2013, Ryu23},degeneracies~\cite{Nussinov}, conserved quantities~\cite{SymmetryConservedQuantity},  decoherence-free subspaces, and noiseless subsystems~\cite{DecoherenceFree}, crucial for condensed matter physics and quantum information science. 

Symmetry affects not only the equilibrium or steady-state properties of the system, but also the speed and dissipative properties in finite-time and nonequilibrium processes. 
When we consider permutation-invariant $N$
identical two-level systems, the decay rate can be enhanced by a factor of $N$ via collective system-bath coupling effects, termed superradiance~\cite{Dicke, Dicke_review}. 
This example implies that symmetry is strongly connected to the notion of collective advantages, which have been extensively studied in the context of quantum thermodynamics, including setups such as heat engines~\cite{gelbwaser2015power, hardal2015superradiant, Niedenzu_2018, PhysRevLett.124.210603, Tajima21, PhysRevLett.128.180602, Brandner23, kim2022photonic}, quantum batteries~\cite{Binder_2015, PhysRevLett.118.150601, PhysRevLett.128.140501}, information erasure protocols~\cite{Rolandi, liu2024}, and photocells~\cite{PhysRevLett.111.253601}. 
It is therefore expected that designing quantum devices that respect symmetry leads to the suppression of unwanted energetic costs, crucial for charge transport dynamics and quantum information processing protocols. 
However, a general framework that addresses the influence of symmetry in finite-time and nonequilibrium thermodynamic processes has not been well established. 

\begin{figure}[t]
    \centering
    \includegraphics[width=0.65\linewidth]{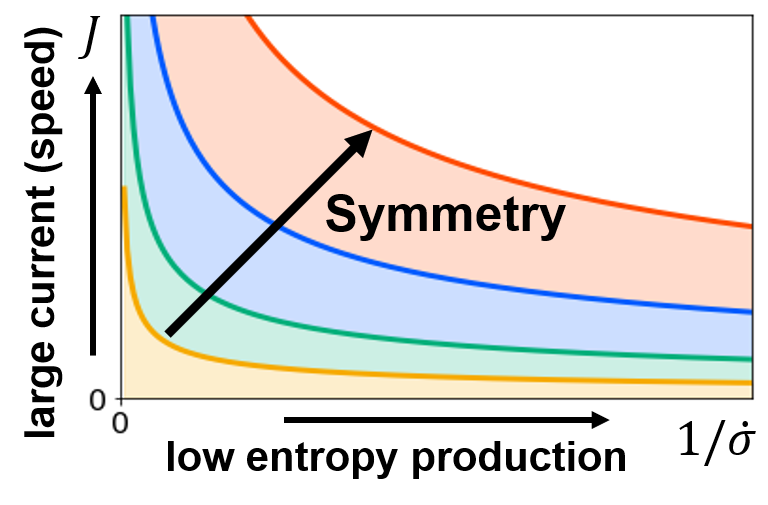}
    \caption{Schematic picture of the current-dissipation trade-off relation $2J^{2}/\dot{\sigma}\leq A$~\cite{Tajima21}, where $J$ is the heat current and $\dot{\sigma}$ is the entropy production rate. When the open system dynamics respects certain type of symmetry, the upper bound $A$ can be enhanced (see Fig.~\ref{fig1})), 
    \revision{and thus higher values of the ratio $J^2/\dot{\sigma}$ can be realized.}
    We derive the fundamental limit of this enhancement [see Eq.~(\ref{generalAbound})].  }
    \label{fig0}
\end{figure}

To develop a general theory to describe the impact of symmetry in finite-time thermodynamics, we pay attention to the thermodynamic speed limit inequalities~\cite{Funo19, PhysRevX.13.011013, PhysRevLett.126.010601} and trade-off relations~\cite{Shiraishi19, Tajima21}, which set generic upper bounds on the speed of state transformation and the change of the expectation values of physical quantities in open quantum systems.  
\revision{These relations indicate that increasing the average jump rate allows having smaller energetic costs (entropy production) while fixing the duration of the process (see Fig.~\ref{fig0}).}
%{\color{red} These relations indicate that increasing the average jump rate allows suppressing the entropy production (the energetic cost) while operating at high speed (see Fig.~\ref{fig0}).}  
Therefore, investigation of symmetry in thermodynamic trade-off relations provides a unified approach to understanding collective advantages in quantum thermodynamics. 
Moreover, in view of the close relation between degeneracy and symmetry~\cite{Nussinov, DecoherenceFree}, such investigation allows symmetry-based understanding of the effect of degeneracy and coherence on quantum thermodynamics~\cite{Tajima21}. 

In this Letter, we develop a generic framework describing the fundamental limit of symmetry-based enhancement in finite-time thermodynamics (see Fig.~\ref{fig0}). 
Specifically, we derive a general upper bound on the average jump rate, showing that the number of degeneracy sets the maximum enhancement. We also derive the symmetry condition on the quantum state and the jump operators that saturates the obtained bound. 
As an application, we consider a permutation-invariant $N$ two-level systems, and discuss the scaling behavior of the power and efficiency of heat engines, based on the power-efficiency trade-off relation~\cite{Tajima21}. 
The obtained theory predicts the possibility of realizing a heat engine that operates near the Carnot efficiency as $\eta=\eta_{\mathrm{Car}}-O(1/N)$ while producing the power that scales from $O(N^{2})$ to exponential, which goes beyond the scaling realized by the superradiant heat engine models~\cite{hardal2015superradiant, Niedenzu_2018}.

%%%%%%%%%%%%%%%%%%%%%%%%%%%%%%%%%%%%%%%%%%%

\textit{Setup.---} 
We assume that the system of interest is interacting with a heat bath. The interaction Hamiltonian reads $H_{\rm int}= \sum_{a}S_{a}\otimes B_{a}$, where $S_{a}$ and $B_{a}$ are Hermitian operators of the system and the bath, respectively. 
We describe the reduced dynamics of the system by considering a weak-coupling and Born-Markov-Secular approximations. Following the standard derivation under these conditions, we obtain the GKSL form of the master equation~\cite{GKS, Lindblad, Breuer} (we take $\hbar=1$)
\beq
\partial_{t}\rho = \mathcal{L}\rho =  -i[H,\rho] + \sum_{a,\omega}\gamma_{a,\omega} \mathcal{D}[L_{a,\omega}] \rho , \label{ME} 
\eeq
where $\mathcal{D}[L]\rho = L \rho L^{\dagger} - (1/2)\{ L^{\dagger}L,\rho \}$ is the dissipator, $\gamma_{a,\omega}$ is the decay rate, and $L_{a,\omega}=\sum_{\omega=\epsilon_{k}-\epsilon_{l}}\Pi_{l}S_{a}\Pi_{k}$ is the Lindblad jump operator that let the system jump from one energy eigenstate to another with their energy difference equal to $\omega$. Here, $\Pi_{k}$ is the projection to the $k$-th energy eigenspace $\mathcal{S}_{k}$, and $\epsilon_{k}$ is the $k$-th energy eigenvalue, i.e., $H=\sum_{k}\epsilon_{k}\Pi_{k}$. We denote $\mathcal{N}_{k}=\dim(\Pi_{k})$ as the number of degeneracy of the $k$-th energy. 
We further assume the detailed balance condition $\gamma_{a,-\omega}=\gamma_{a,\omega}e^{-\beta \omega}$ to make Eq.~({\ref{ME}) thermodynamically consistent, where $\beta$ is the inverse temperature of the heat bath.

One of the main objectives in the field of stochastic thermodynamics is to minimize the entropy production in finite-time processes~\cite{Deffner_2020, PhysRevLett.124.110606, Esposito10, OptimalTransport, Kosloff_review}.
\revision{To this end, we focus on the activity~\cite{Funo19, PhysRevLett.121.070601} $A_{\rm act}=\sum_{\omega}A_{\omega}$ and an activity-like quantity $A=\sum_{\omega}\omega^{2}A_{\omega}$, where
\beq
A_{\omega}(\rho,L_{a,\omega})=\sum_{a}\gamma_{a,\omega}\Tr[ L_{a,\omega}^{\dagger}L_{a,\omega}\rho], \label{jumprate}
\eeq
is the average jump rate with fixed transition energy $\omega$. 
Note that these quantities set a time-scale of the system and play a fundamental role in stochastic thermodynamics. Specifically, the current-dissipation trade-off relation reads $2J^{2}/\dot{\sigma} \leq A$, where $J$ is the heat current, $\dot{\sigma}$ is the entropy production rate~\cite{PhysRevLett.117.190601, Shiraishi19, Tajima21}.}
This trade-off bound is achievable in specific examples \cite{Tajima21}. Therefore, suppressing the entropy production while producing a large heat current becomes possible when the value of $A$ is increased (see Fig.~\ref{fig0}).

Moreover, $A_{\omega}$ quantifies the transition rate from a $k$-th energy eigenstate $|\psi_{k}\rangle$ to a $l$-th energy eigenstate $|\psi_{l}\rangle$, given by $\sum_{a,\omega}\gamma_{a,\omega}|\langle \psi_{l}|L_{a,\omega}|\psi_{k}\rangle|^{2}=A_{\epsilon_{k}-\epsilon_{l}}(|\psi_{k}\rangle\langle \psi_{k}|)$. 
Therefore, having a large $A_{\omega}$ allows increasing the emission of photons to the environment, with the possibility of realizing the enhancement that goes beyond superradiance.

%\textit{Symmetry and degeneracy.---} 
\textit{Symmetry.---} 
In what follows, we develop a symmetry-based theory that quantifies the limit of the enhancement of $A_{\omega}$.
To this end, we introduce a symmetry group $G$ and assume that the Hamiltonian $H$ is invariant under this group: $[H,V_{g}]=0$ for all $g\in G$\revision{, where $V_{g}$ is a unitary representation of the group.}
We then classify quantum states and jump operators based on $V_{g}$, which constitutes the core of our analysis.
\revision{To this end, we assume in the main text that we take appropriate $G$ and $V_{g}$ whose precise condition is given in Appendix A, such that the symmetry represented by $V_{g}$ perfectly characterizes the structure of the energy eigenspaces of $H$. Note that the examples that we discuss later satisfy this condition. On the other hand, the above condition is not satisfied when $G$ does not represent all the symmetry of the Hamiltonian $H$, e.g., by choosing $G$ as a trivial group, or by only choosing $G$ as the permutation group for the permutation and even number bit-flip invariant system considered in {\it Example 2}.
In Appendix B, we generalize our results to any choice of $G$ and $V_{g}$. In particular, the main result~(\ref{generalAbound}) is still valid, but the bound is no longer achievable in general; we thus further derive an achievable bound on $A_{\omega}$.
}

\revision{We point out that $A_{\omega}(\rho,L_{a,\omega}) = \sum_{k}p_{k}A_{\omega}(\rho_{k},L_{a,\omega})$, where $p_{k}=\Tr[\Pi_{k}\rho]$, and $\rho_{k}=\Pi_{k}\rho\Pi_{k}/p_{k}$. 
This relation motivates us to characterize quantum states $\rho$ by their properties of $\rho_{k}$ acting on $\mathcal{S}_{k}$. We now introduce following two special classes of quantum states based on $V_{g}$:}

\begin{itemize}
    \item \textit{Local states} $\rho_{k}^{\rm loc}$, defined by 
%       \item \textit{Local states} $\rho^{\rm loc}$, defined by 
\beq
\frac{1}{|G|}\sum_{g\in G}V_{g}\rho_{k}^{\rm loc}V_{g}^{\dagger}=\frac{1}{\mathcal{N}_{k}} \Pi_{k} . \label{localBlock}
\eeq
This definition means that by randomly mixing a local state by unitary operators $V_{g}$, it gets completely mixed and becomes the maximally mixed state in $\mathcal{S}_{k}$. 
We also introduce the local energy eigenbasis $\mathcal{B}^{\rm loc}_{k}=\{ |\psi_{k}^{\rm loc}(\alpha)\rangle\}_{\alpha=1}^{\mathcal{N}_{k}}$ for $\mathcal{S}_{k}$, where each element of $\mathcal{B}^{\rm loc}_{k}$ satisfies Eq.~(\ref{localBlock}).
%We remark that this is a natural generalization of the local states of the identical two-level systems. 
%Indeed, as we discuss in the example section, we have $|e\rangle^{\otimes k}\otimes |g\rangle^{\otimes N-k} \in \mathcal{B}^{\rm loc}_{k}$, representing $k$ ``local" excitations of individual two-level systems. With this in mind, we also call $\rho_{k}^{\rm loc}$ as the local states in generic situations. 

    %\item \textit{Symmetric state} $|\psi^{\rm sym}_{k}\rangle$, defined by  
    %\beq
    %V_{g}|\psi_{k}^{\rm sym}\rangle= |\psi_{k}^{\rm sym}\rangle\ \text{ for all } g. \label{defSymState}
    %\eeq
    
    \item \revision{\textit{Symmetric states} $\rho^{\rm sym}_{k}$, defined by  
    \beq
    V_{g}\rho_{k}^{\rm sym}= \rho_{k}^{\rm sym}V_{g}^{\dagger} = \rho_{k}^{\rm sym} \ \text{ for all } g. \label{defSymState}
    \eeq
%    where $\rho^{\rm sym}_{k}=\Pi_{k}\rho^{\rm sym}\Pi_{k}/p_{k}$ and $p_{k}=\Tr[\Pi_{k}\rho^{\rm sym}]$. 
    This definition means that symmetric states do not change by the action of $V_{g}$.  
    }
    
    %\revision{As shown later, the symmetric state achieves the upper bound of the enhancement of $A_\omega$. However, unlike the local state, the symmetric state does not necessarily always exist. The existence of the symmetric state requires the condition $\dim(\mathcal{K}_{j}^{k})=1$ for one $j$. And when $\dim(\mathcal{H}_{j}^{k})=1$, symmetric state is unique and can be written as $\rho^{\rm sym}_{k}=|\psi_{k}^{\rm sym}\rangle\langle \psi_{k}^{\rm sym}|$, where $|\psi_{k}^{\rm sym}\rangle$ is defined by $V_{g}|\psi_{k}^{\rm sym}\rangle = |\psi_{k}^{\rm sym}\rangle$ for all $g$. {\color{red}Unlike $\dim(\mathcal{H}_{j}^{k})=1$, it is not always possible to take $G$ and $V_g$ to satisfy the condition $\dim(\mathcal{K}_{j}^{k})=1$.} However, the examples we discuss later satisfy this condition. Therefore, we assume this condition throughout this paper. {\color{red}Tajima-comment: Maybe we can omit this assumption, and rewrite the main results related to the symmetric states as the forms of ``if the symmetric state exists, then..."}
    %}} %Similar to the definition of local states, this definition is a natural generalization of the symmetric Dicke states of the identical two-level systems.
\end{itemize}

Next, we consider the symmetry-based classification of jump operators. To this end, we introduce the following covariant condition (so-called weak symmetry condition) for the Liouvillian $\mathcal{L}(V_{g}xV^{\dagger}_{g})=V_{g}\mathcal{L}(x)V^{\dagger}_{g}$ for any operator $x$ and for all $g\in G$~\cite{Buca12}. This condition imposes the Liouvillian $\mathcal{L}$ to preserve symmetry. 
Now, we introduce two special types of the jump operators:\revision{  
\begin{itemize}
    \item \textit{Local jump operators} $\{L^{\rm loc}_{a,\omega}\}$, defined by 
    \beq
    \left[ (L^{\rm loc}_{a,\omega})^{\dagger}L^{\rm loc}_{a,\omega} , |\psi_{k}^{\rm loc}(\alpha)\rangle \langle \psi_{k}^{\rm loc}(\alpha)| \right] = 0 \text{ for all } \alpha.  \label{LlocBlock}
    \eeq 
    Therefore, local jump operators do not create coherence between local states $|\psi_{k}^{\rm loc}(\alpha)\rangle$ and $|\psi_{k}^{\rm loc}(\alpha')\rangle$. 

    \item \textit{Symmetric jump operators} $\{L^{\rm sym}_{a,\omega}\}$, defined by 
    \beq
    V_{g}L^{\rm sym}_{a,\omega}=L^{\rm sym}_{a,\omega}V^{\dagger}_{g} = L^{\rm sym}_{a,\omega} \ \text{ for all } g . \label{LsymV}
    \eeq 
    Similar to symmetric states, symmetric jump operator do not change by the action of $V_{g}$. 
\end{itemize}
}

\revision{
To see why we call $\{L^{\rm loc}_{a,\omega}\}$ and $\{ |\psi_{k}^{\rm loc}(\alpha)\rangle\}$ as local, let us consider a permutation-invariant, $N$ identical two-level systems discussed in {\it Example 1}. We then find that $|\psi_{k}^{\rm loc}\rangle=|e\rangle^{\otimes k}\otimes |g\rangle^{\otimes N-k} \in \mathcal{B}^{\rm loc}_{k}$, where $|g\rangle$ and $|e\rangle$ denote the ground and excited states of individual two-level systems. This state $|\psi_{k}^{\rm loc}\rangle$ is ``local" in the sense that it is a tensor product of individual two-level states, and does not have superpositions among different subsystems. We also note that the jump operators $\{\sigma_{i}^{-}\}_{i=1}^{N}$ satisfy Eq.~(\ref{LlocBlock}), where $\sigma_{i}^{-}$ is the lowering operator that acts ``locally" on the $i$-th subsystem. With these in mind, we also call $\rho_{k}^{\rm loc}$ and $L_{a,\omega}^{\rm loc}$ as local quantum states and local jump operators in generic situations.  
}

In what follows, we utilize the above classification of the quantum states and jump operators and derive general properties of $A_{\omega}$, including its upper bound. 

\begin{table}
    \centering
    \begin{tabular}{c|c|c} 
             & local jump op. & sym. jump op. \\
         & $\{L_{a,\omega}^{\rm loc}\}$ & $\{L^{\rm sym}_{a,\omega}\} $  \\   \hline 
       local state  & $c_{k}(L_{a,\omega}^{\rm loc})$  & $c_{k}(L_{a,\omega}^{\rm sym})$  \\
       $\rho_{k}^{\rm loc}$ & & (no enhancement) \\ \hline
        sym. state & $c_{k}(L_{a,\omega}^{\rm loc})$ & $\mathcal{N}_{k}c_{k}(L_{a,\omega}^{\rm sym})$  \\ 
        $\rho_{k}^{\rm sym}$ & (no enhancement) & (max. enhancement) \\
        \hline
    \end{tabular}
    \caption{\revision{Classification of the enhancement of $A_{\omega}$. If either the state or the jump operator is local, $A_{\omega}$ is given by $c_{k}$, and cannot be enhanced~(\ref{Alocstate}) and (\ref{Alocjump}). If both the state and the jump operator are symmetric, $A_{\omega}$ is maximally enhanced, characterized by the number of degeneracy $\mathcal{N}_{k}$~(\ref{generalAbound}).}
    }
    \label{tab1}
\end{table}

\textit{No enhancement condition.---} 
\revision{First, we show in the supplemental material~\cite{supplement} that
\beqa
A_{\omega}(\rho^{\rm loc},\{L_{a,\omega}\})&=&\sum_{k}p_{k}c_{k}(L_{a,\omega}),  \label{Alocstate} \\ 
A_{\omega}(\rho,\{L_{a,\omega}^{\rm loc}\}) &=& \sum_{k}p_{k}c_{k}(L_{a,\omega}^{\rm loc}), \label{Alocjump}
\eeqa
where $c_{k}(L_{a,\omega})=\mathcal{N}_{k}^{-1}\sum_{a}\gamma_{a,\omega}\Tr[\Pi_{k}L_{a,\omega}^{\dagger}L_{a,\omega}\Pi_{k}]$ is the square of the Hilbert-Schmidt norm of the jump operators acting on the subspace $\mathcal{S}_{k}$ divided by its dimension, and $\rho^{\rm loc}=\sum_{k}p_{k}\rho_{k}^{\rm loc}$.
%$p_{k}=\Tr[\rho^{\rm loc}\Pi_{k}]$ and $\rho_{k}^{\rm loc}=\Pi_{k}\rho^{\rm loc}\Pi_{k}/p_{k}$ in Eq.~(\ref{Alocstate}). 
Note that if we consider a trivial rescaling $\sqrt{\gamma_{a,\omega}}L_{a,\omega}\rightarrow \sqrt{C\gamma_{a,\omega}}L_{a,\omega}$, the average jump rate is rescaled as $A_{\omega}\rightarrow CA_{\omega}$, where $C$ is a constant. Therefore, it would be reasonable to analyze the amount of $A_{\omega}$ in units of some norm of the jump operators, and we have therefore introduced $c_{k}(L_{a,\omega})$. 
Equations~(\ref{Alocstate}) and (\ref{Alocjump}) show that the norm of the jump operators $c_{k}(L_{a,\omega})$ sets the value of $A_{\omega}$ if at least one of the state and jump operator is local. 
}

\textit{Maximum enhancement condition.--- } 
We now analyze, to what extent $A_{\omega}$ can be enhanced. In the supplemental material~\cite{supplement}, we show a general upper bound on $A_{\omega}$, for any density matrix $\rho$ and jump operators $\{L_{a,\omega}\}$, expressed as
\revision{
\beq
A_{\omega}(\rho,\{L_{a,\omega}\}) \leq  \sum_{k}p_{k}\mathcal{N}_{k}c_{k}(L_{a,\omega}) , \label{generalAbound}
\eeq
showing that $A_{\omega}$ can be enhanced up to $\mathcal{N}_{k}$ times the norm of jump operators $c_{k}$ for each $k$-th subspace. The equality condition in~(\ref{generalAbound}) is achieved by a combination of symmetric states and jump operators, given by 
\beq
A_{\omega}(\rho^{\rm sym},\{L_{a,\omega}^{\rm sym}\})=\sum_{k}p_{k}\mathcal{N}_{k}c_{k}(L_{a,\omega}^{\rm sym}), \label{Asym}
\eeq
where $\rho^{\rm sym}=\sum_{k}p_{k}\rho_{k}^{\rm sym}$. 
See Tab.~\ref{tab1} for the summary of the scaling of $A_{\omega}$ for different states and jump operators.} 

In what follows, we show specific examples and construct jump operators that realize better scaling of $A_{\omega}$ compared to the superradiance model. We also show that such jump operators allow enhancing the output power and efficiency of heat engines.

\begin{figure}
    \centering
    \includegraphics[width=0.95\linewidth]{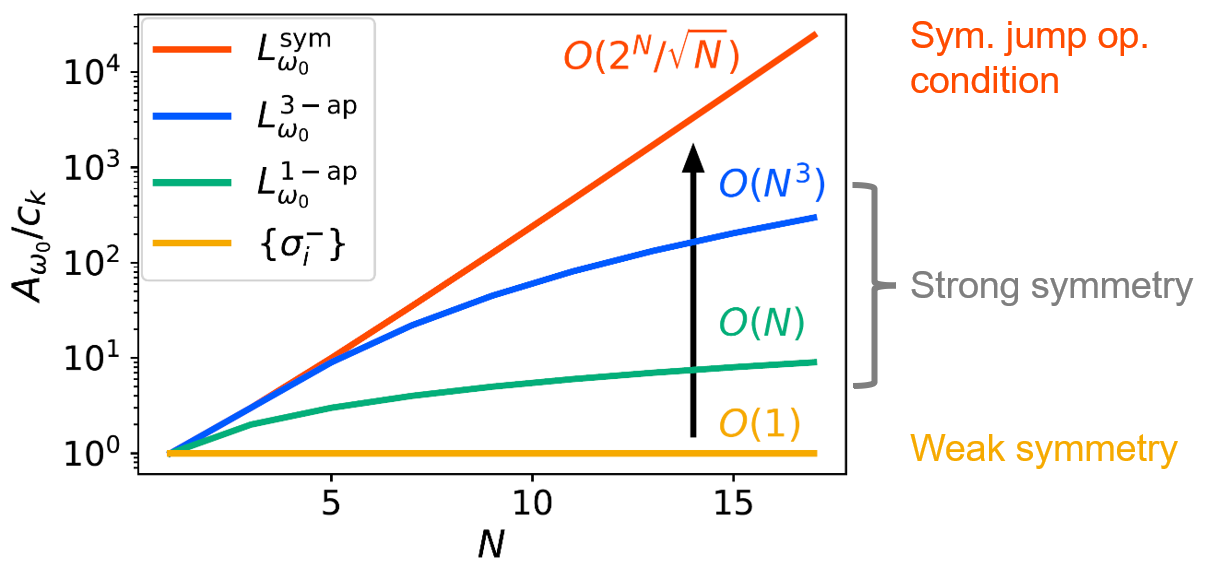}
    \caption{\revision{Plot of $A_{\omega_{0}}(\rho_{k}^{\rm sym},L_{a,\omega_{0}})/c_{k}$ for a permutation-invariant $N$ two-level systems model. We plot the case $k=\lceil N/2\rceil$, i.e., half of the two-level systems are excited. The red curve is obtained by using the symmetric jump operator $L^{\rm sym}_{\omega_{0}}$, demonstrating the optimal enhancement of $A_{\omega_{0}}$. 
    The blue curve is obtained by using $L^{3\text{-ap}}_{\omega_{0}}$, showing $O(N^{3})$ scaling. The green curve is obtained by using the conventional collective jump operator $L^{1\text{-ap}}_{\omega_{0}}=\sum_{i}\sigma^{-}_{i}$ in the analysis of super-radiance, showing $O(N)$ scaling. The orange line is obtained by using local jump operators $\{\sigma^{-}_{i}\}$, demonstrating $A_{\omega_{0}}/c_{k}=1$.} 
    }
    \label{fig1}
\end{figure}

\textit{Eample 1: permutation invariance.---} 
We now apply our results to a permutation-invariant model. Let the system Hamiltonian be $N$ identical two-level systems $H=(\omega_{0}/2)\sum_{i=1}^{N}\sigma^{z}_{i}$, where $\sigma^{z}_{i}$ is the $z$-component of the Pauli matrix for the $i$-th system. This Hamiltonian is invariant under interchange of subsystem labels $i$, and is thus invariant under the permutation group $S_{N}$. One example of the local state is given by $|\psi_{k}^{\rm loc}\rangle =|e\rangle^{\otimes k}\otimes |g\rangle^{\otimes N-k}$, representing the state with $k$ ``local" excitations of the two-level systems. 
On the other hand, the symmetric state $\rho_{k}^{\rm sym}=|\psi_{k}^{\rm sym}\rangle\langle \psi_{k}^{\rm sym}|$ is given by the symmetric Dicke state $|\psi_{k}^{\rm sym}\rangle = \mathcal{N}_{k}^{-1/2}\sum_{g}V_{g}|\psi_{k}^{\rm loc}\rangle$, where $\mathcal{N}_{k}={}_{N}C_{k}$ is the number of degeneracy. 

In what follows, we demonstrate the obtained results by explicitly calculating $A_{\omega_{0}}(\rho_{k}^{\rm sym},L_{a,\omega_{0}})$ for different jump operators $L_{a,\omega_{0}}$ that \revision{removes} one excitation from the system. \revision{For mixed states $\rho^{\rm sym}=\sum_{k}p_{k}\rho_{k}^{\rm sym}$, the scaling of $A_{\omega}$ is simply given by the linear combination $\sum_{k}p_{k}A_{\omega_{0}}(\rho_{k}^{\rm sym},L_{a,\omega_{0}})$.}  
Note that the case of adding one excitation to the system ($\omega=-\omega_{0}$) can be similarly obtained by using the jump operators $L_{a,-\omega_{0}}=L_{a,\omega_{0}}^{\dagger}$. In the following, we set $\gamma_{a,\omega_{0}}=\gamma_{\downarrow}$.  

The symmetric jump operator~(\ref{LsymV}) is given by 
\beq
L^{\rm sym}_{\omega_{0}}= \sum_{m=0}^{\lceil N/2\rceil-1}L^{(m)}_{\omega_{0}}   , \label{SR-symL}
\eeq
where $L^{(m)}_{\omega_{0}}=\sum \sigma_{i_{1}}^{-}\cdots \sigma_{i_{m+1}}^{-}\sigma_{l_{1}}^{+}\cdots \sigma_{l_{m}}^{+}$ and the summation is taken over $(i_{1}<\cdots<i_{m+1})\neq (l_{1}<\cdots <l_{m})$, and $\lceil\bullet\rceil$ is the ceiling function. We note that $L^{(m)}_{\omega_{0}}$ includes all possible combinations of $2m+1$-body jump operators that remove one excitation from the system. 
Using Eq.~(\ref{SR-symL}), we find that $A_{\omega_{0}}(\rho_{k}^{\rm sym},L^{\rm sym}_{\omega_{0}}) =  \mathcal{N}_{k} c_{k}$, with $c_{k}={}_{N}C_{k-1}\gamma_{\downarrow}$. When $k= \lceil N/2\rceil$, we use Stirling's formula and obtain $\mathcal{N}_{N/2}\sim \sqrt{2/(\pi N)}2^{N}$, showing an exponential scaling (see also Fig.~\ref{fig1}). 

\revision{Note that Eq.~(\ref{SR-symL}) consists of many-body system operators, which makes it challenging to realize the above optimal scaling in practice.
In the following, we therefore approximate Eq.~(\ref{SR-symL}) by taking the first $2n+1$-body terms $L^{2n+1\text{-ap}}_{\omega_{0}}=\sum_{m=0}^{n}L^{(m)}_{\omega_{0}}$ and analyze the scaling of the average jump rate.} 
In particular, the 1-body approximation reproduces the collective jump operator $L^{1\text{-ap}}_{\omega_{0}}=\sum_{i=1}^{N}\sigma_{i}^{-}$, which is used in the study of superradiance~\cite{Dicke}. The jump operator $L^{1\text{-ap}}_{\omega_{0}}$ satisfies strong symmetry~\cite{Buca12} $[L^{1\text{-ap}}_{\omega_{0}},V_{g}]=0$, but does not satisfy Eq.~(\ref{LsymV}). The scaling reads $A_{\omega_{0}}(\rho_{k}^{\rm sym},L^{1\text{-ap}}_{\omega_{0}}) = (N-k+1)c_{k}$ with $c_{k}=k\gamma_{\downarrow}$. 
By considering the next order term $L^{3\text{-ap}}_{\omega_{0}}$, we find that $A_{\omega_{0}}(\rho_{k}^{\rm sym},L^{3\text{-ap}}_{\omega_{0}})
=(N-k+1)[1+(N-k)(k-1)/2]c_{k}$, with $c_{k}=k(1+(N-k)(k-1)(k-2)/2)\gamma_{\downarrow}$. When $k = \lceil N/2\rceil$, $A_{\omega_{0}}/c_{k}$ scales $O(N^{3})$, compared to the case of $O(N)$ scaling for the conventional superradiance model (see Fig.~\ref{fig1}). 

Finally, we consider local jump operators $\{\sigma_{i}^{-}\}_{i=1}^{N}$. This set of jump operators satisfies the weak symmetry, but does not satisfy the strong symmetry nor Eq.~(\ref{LsymV}). We find that $A_{\omega_{0}} =  c_{k}=k\gamma_{\downarrow}$, consistent with Eq.~(\ref{Alocjump}). 

\begin{figure}
    \centering
    \includegraphics[width=0.99\linewidth]{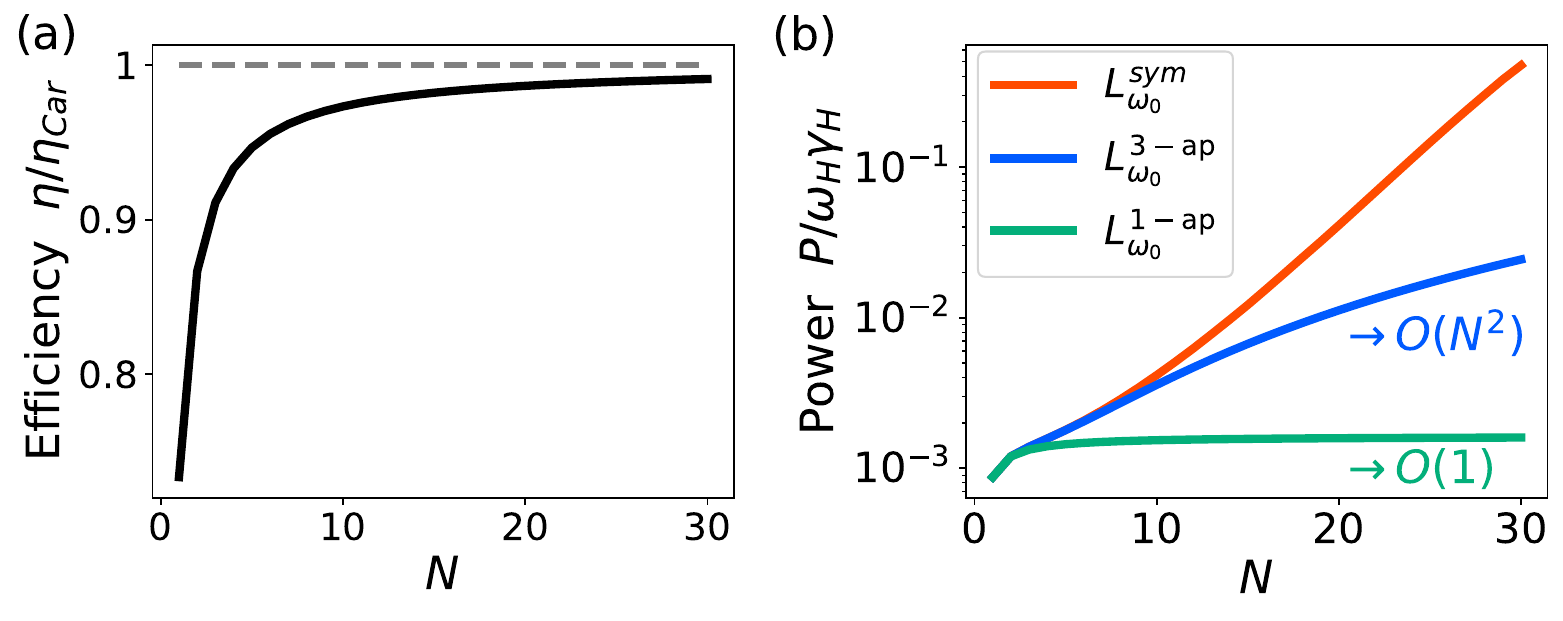}
    \caption{ Plot of the efficiency $\eta/\eta_{\rm Car}$ and the power $P/\omega_{H}\gamma_{H}$ of the permutation invariant $N$ two-level system heat engine. (a) The efficiency (black curve) scales as $\eta=\eta_{\rm Car}-b/N$ and asymptotically reaches the Carnot efficiency (gray dashed line). 
    (b) By using the conventional superradiance model ($L^{1\text{-ap}}_{\omega_{0}}$), the power saturates at large $N$ (green curve). If we use $L^{3\text{-ap}}_{\omega_{0}}$, the power scales $O(N^{2})$ (blue curve)\revision{, and if we use $L_{\rm sym}$, the power scales exponentially (red curve). Here, $\omega_{\rm H}$ and $\gamma_{\rm H}$ are the energy level splitting and the decay rate of the system during the hot thermalization stroke, respectively.} %See the supplementary material~\cite{supplement} for details.
    }
    \label{fig2}
\end{figure}

\textit{Application of Example 1 to heat engines.---} 
We apply the permutation-invariant model to the analysis of heat engines. The output power $P$ and the heat-to-work conversion efficiency $\eta$ of the heat engines satisfy the power-efficiency trade-off relation~\cite{Shiraishi19, Tajima21} 
\beq
\frac{P}{\eta_{\rm Car}-\eta} \leq c\bar{A}, \label{PEtradeoff}
\eeq
where $\eta_{\rm Car}=1-\beta_{H}/\beta_{C}$ is the Carnot efficiency; $\beta_{H}$ and $\beta_{C}$ are the inverse temperatures of the hot and cold baths; $c=\beta_{C}\eta_{\rm Car}/[2(2-\eta_{\rm Car})^{2}]$ is a constant; and \revision{$\bar{A}=\tau^{-1}\int^{\tau}_{0}dt\sum_{\omega}\omega^{2}A_{\omega}$}, where $\tau$ is the duration of time to complete one engine cycle. 
%{\color{red}Not needed?} Because the population $p_{k}$ of higher energy level becomes smaller, the scaling condition $k=\lceil N/2\rceil$ plotted in Fig.~\ref{fig1} cannot be achieved in general. Nevertheless, we show that the above analysis can be utilized to design a heat engine with high power and efficiency as follows. 
%typically follows an exponential distribution $\ln p_{k}\propto -k$ during the heat engine operation, 
%the contribution from $k=[N/2]$ plotted in Fig.~\ref{fig1} becomes negligibly small in large $N$

\revision{We consider a finite-time quantum Otto heat engine using the jump operators $L^{1\text{-ap}}_{\pm\omega_{0}}$, $L^{3\text{-ap}}_{\pm\omega_{0}}$, and $L_{\pm\omega_{0}}^{\rm sym}$ (see the supplementary material~\cite{supplement} for details). Because of the strong symmetry $[H,V_{g}]=[L_{a,\omega},V_{g}]=0$, %the density matrix of the system can be block-decomposed into subspaces which evolve independently under $\mathcal{L}$~\cite{Buca12, SymmetryConservedQuantity}. Specifically, 
if the initial state is prepared by $\rho^{\rm sym}(0)=\sum_{k}p_{k}(0)\rho_{k}^{\rm sym}$, the density matrix at later times remains in the same symmetric Dicke subspace spanned by $|\psi_{k}^{\rm sym}\rangle$, and generically takes the form $\rho(t)=\sum_{k}p_{k}(t)\rho_{k}^{\rm sym}$. Therefore, the scaling of $A_{\omega}$ discussed in the previous section can be directly applied to investigate the scalings of power and efficiency as follows (see also Appendix C).}

We choose a protocol such that the efficiency asymptotically reaches the Carnot efficiency as $\eta=\eta_{\rm Car}-b/N$, where $b$ is a constant (see Fig.~\ref{fig2} (a)). 
In Fig.~\ref{fig2} (b), we show a numerical plot of the output power. The green curve shows the superradiant heat engine setup~\cite{hardal2015superradiant}, where the jump operators are given by $L^{1\text{-ap}}_{\pm\omega_{0}}$.  
An analytical calculation shows $\bar{A}=O(N)$, and combined with Eq.~(\ref{PEtradeoff}), the power is expected to scale $O(1)$, which is consistent with the numerical plot. 
If we instead consider $L^{3\text{-ap}}_{\pm \omega_{0}}$, a similar analysis implies that the power scales $O(N^{2})$.
\revision{By further considering $L^{\rm sym}_{\pm\omega_{0}}$, we find an exponential scaling of the power (see Fig.~\ref{fig2}). }

\textit{Example 2: permutation and even bit flip invariance.---} 
We next apply our results to a permutation and even number bit-flip (e.g., $V_{g}=\sigma_{i}^{x}\sigma_{j}^{x}$) invariant system. The Hamiltonian reads $H = \epsilon \prod_{i=1}^{N}\sigma_{i}^{z}$. When $N=4$, this Hamiltonian appears in the Kitaev's toric code model as one of the stabilizer operators~\cite{KITAEV20032}. The eigenenergies and the number of degeneracies are given by $\pm \epsilon$ and $\mathcal{N}_{\pm \epsilon}=2^{N-1}$, respectively. 
The symmetric state reads $|\psi^{\rm sym}_{\pm\epsilon}\rangle = ( |+\rangle^{\otimes N} \pm |-\rangle^{\otimes N} )/\sqrt{2}$, where $|\pm\rangle=(|e\rangle\pm |g\rangle)/\sqrt{2}$. 
The symmetric jump operator reads 
\beq
L^{\rm sym}_{2\epsilon}= \sum_{m=0}^{[N/2]-1}\sum_{i_{1}<\cdots<i_{2m+1}} \Pi_{-\epsilon}\sigma^{x}_{i_{1}}\cdots \sigma^{x}_{i_{2m+1}}\Pi_{\epsilon}. \label{ToricLsym}
\eeq
Using Eq.~(\ref{ToricLsym}), we obtain $A_{2\epsilon}(|\psi^{\rm sym}_{\epsilon}\rangle\langle\psi^{\rm sym}_{\epsilon}|,L^{\rm sym}_{2\epsilon})= \mathcal{N}_{\epsilon}c_{\epsilon}$ with $c_{\epsilon}=\mathcal{N}_{\epsilon}\gamma_{\downarrow}$, consistent with Eq.~(\ref{Asym}). This scaling behavior allows us to construct a heat engine model that achieves $\eta_{\rm Car}-\eta=O(1/\mathcal{N}_{\epsilon})$ and $P=O(\mathcal{N}_{\epsilon})$ discussed in Ref.~\cite{Tajima21}.

\textit{Conclusion.---} 
We have shown that the number of degeneracy sets a general upper bound on the average jump rate, and derived a symmetry condition on the quantum states and jump operators that saturates the obtained bound.
%\revision{We have introduced two special classes of quantum states and jump operators based on the symmetry of the model and classified the enhancement of the average jump rate.} 
The obtained results clarify the effect of symmetry in finite-time thermodynamic trade-off relations.
As an application, we consider a quantum heat engine composed of permutation invariant $N$ two-level systems. In contrast to the conventional super-radiant heat engine model~\cite{hardal2015superradiant, Niedenzu_2018}, \revision{we obtained from $O(N^{2})$ to exponential enhancement of the output power by designing the jump operators that better respects the obtained symmetry condition.} 

\revision{
An interesting future direction is to generalize the obtained framework to generic situations, for example, when the detailed balance is violated~\cite{PhysRevLett.131.040401}, the system dynamics is generically non-Markovian~\cite{RevModPhys.88.021002, RevModPhys.89.015001}, and there are non-reciprocal interactions~\cite{PhysRevX.5.021025, 10.21468/SciPostPhysLectNotes.44}. The theoretical framework developed in this paper is anticipated to lead not only to designing high-performance heat engines but also to realizing fast and energy-efficient information processing devices and charge transport devices.
}

\begin{acknowledgments}
\textit{Acknowledgements.---} We thank Atsushi Noguchi for useful discussions. 
This work was supported by MEXT KAKENHI Grant-in-Aid for Transformative
Research Areas B ``Quantum Energy Innovation” Grant Nos. JP24H00830 and JP24H00831. 
K. F. acknowledges support from JSPS KAKENHI Grant No. JP23K13036 and JST ERATO Grant No. JPMJER2302, Japan. 
H.T. was supported by JST PRESTO No. JPMJPR2014, JST MOONSHOT No. JPMJMS2061.
\end{acknowledgments}

\appendix 

\revision{
\textbf{Appendix A: Details on the choice of $G$ and $V_{g}$ assumed in the main text}

In this appendix, we show technical details on the appropriate choice of $G$ and $V_{g}$ assumed in the main text. We again note that the main result~(\ref{generalAbound}) can be generalized to arbitrary $G$ and $V_{g}$, as shown in Appendix B. 

To begin with, we note that the commutation relation $[V_{g},H]=0$ implies $[V_{g},\Pi_{k}]=0$. Then, $V_{g}$ can be decomposed as $V_{g} = \bigoplus_{k} V_{g}^{k}$, where $V_{g}^{k}$ acts on $\mathcal{S}_{k}$. Each $V_{g}^{k}$ is further decomposed into irreducible representations as $V^{k}_{g}=\bigoplus_{j} I_{\mathcal{H}_{j}^{k}}\otimes \mathcal{V}^{k}_{j}(g)$, 
where $\mathcal{S}_{k}$ is decomposed as $\mathcal{S}_{k}=\bigoplus_{j} \mathcal{H}_{j}^{k}\otimes \mathcal{K}_{j}^{k}$, $\mathcal{V}_{j}^{k}(g)$ is an irreducible representation of $G$ acting on the subspace $\mathcal{K}_{j}^{k}$, $I_{\mathcal{H}_{j}^{k}}$ is the identity matrix acting on the subspace $\mathcal{H}_{j}^{k}$, and $j$ labels irreducible representations~\cite{Brandner23, goodman2009symmetry, etingof2011introductionrepresentationtheory}. Here, a representation $\mathcal{V}_{j}^{k}(g)$ of $G$ acting on $\mathcal{K}_{j}^{k}$ is called irreducible if $\mathcal{K}_{j}^{k}$ has no nontrivial subspace that is invariant under the action of $\mathcal{V}_{j}^{k}(g)$ for all $g$. Therefore, the subspace $\mathcal{H}_{j}^{k}$ is invariant under the operation of $V^{k}_{g}$, and $\mathcal{K}_{j}^{k}$ is the only subspace in which $V_{g}^{k}$ nontrivially acts on.
When there exist $(j,k)$ such that $\dim(\mathcal{H}_{j}^{k})\geq 2$, the symmetry represented by $V_g$ does not perfectly characterize the structure of the energy eigenspaces of $H$, due to these invariant subspaces.
Fortunately, for a given $H$, we can always take appropriate $G$ and $V_g$ that satisfies $\dim(\mathcal{H}_{j}^{k})=1$ for any $j$ and $k$ (see the supplementary material~\cite{supplement}).
We also note that the examples we discuss in the main text satisfy the condition $\dim(\mathcal{H}_{j}^{k})=1$ for any $j$ and $k$ for natural $G$ and $V_g$.
Therefore, in the main text, we assume that we take appropriate $G$ and $V_g$ that satisfy $\dim(\mathcal{H}_{j}^{k})=1$ for any $j$ and $k$. 
In Appendix B, we discuss the case $\dim(\mathcal{H}_{j}^{k})\geq 2$ and generalize the main results.
}

\revision{
\textbf{Appendix B: Generalization to arbitrary $G$ and $V_{g}$}

We now show how the main results are generalized to the case of $\dim(\mathcal{H}_{j}^{k})\geq 2$, i.e., arbitrary $G$ and $V_{g}$. To begin with, we introduce operators $\sigma_{j,k}$ and $B_{j,k}^{\omega}$ acting on the subspace $\mathcal{H}_{j}^{k}$ to parametrize quantum states and jump operators as 
\beqa
& &[\rho_{k}]_{\rm inv} = \frac{1}{\sum_{j}\Tr[\sigma_{j,k}]}\bigoplus_{j} \sigma_{j,k} \otimes \frac{I_{\mathcal{K}_{j}^{k}}}{\dim(\mathcal{K}_{j}^{k})},  \label{sup_qstate2} \\
& &\Bigl[\sum_{a}\gamma_{a,\omega}L_{a,\omega}^{\dagger}L_{a,\omega} \Bigr]_{\rm inv} = \bigoplus_{k,j}B_{j,k}^{\omega} \otimes \frac{I_{\mathcal{K}_{j}^{k}}}{\dim(\mathcal{K}_{j}^{k})}, \label{sup_bjump2}
\eeqa
where $[X]_{\rm inv}:=|G|^{-1}\sum_{g\in G}V_{g}XV_{g}^{\dagger}$. We again note that when $\dim(\mathcal{H}_{j}^{k})\geq 2$, $\mathcal{H}_{j}^{k}$ is a nontrivial invariant subspace under the action of $V_{g}$, and the specific form of $\sigma_{j,k}$ and $B_{j,k}^{\omega}$ cannot be constrained based on the symmetry conditions for given $G$ and $V_{g}$. Nevertheless, a general upper bound on the average jump rate can be derived as (see supplementary material~\cite{supplement} for details)
\beqa
A_{\omega}(\rho,\{L_{a,\omega}\}) &\leq& \sum_{k}p_{k}\mathcal{N}_{k}c_{k}(L_{a,\omega}) F(\sigma_{j,k},B_{j,k}^{\omega}) \nonumber \\ 
&\leq &\sum_{k}p_{k}\mathcal{N}_{k}c_{k}(L_{a,\omega}) , \label{app_AgenH2}
\eeqa
where
\beq
F(\sigma_{j,k},B_{j,k}^{\omega})=\frac{\sum_{j}\Tr[\sigma_{j,k}B_{j,k}^{\omega}]}{\sum_{j}\Tr[\sigma_{j,k}]\sum_{j}\Tr[B_{j,k}^{\omega}]} \leq 1, \label{app_AgenH_eq}
\eeq
quantifies the overlap between $\{\sigma_{j,k}\}_{j}$ and $\{B_{j,k}^{\omega}\}_{j}$. 
The obtained relation~(\ref{app_AgenH2}) generalizes the result Eq.~(\ref{generalAbound}) to the case of $\dim(\mathcal{H}_{j}^{k})\geq 2$. It should be noted that the bound~(\ref{generalAbound}) remains valid in this general case; however, the last equality condition in~(\ref{app_AgenH2}) can no longer be characterized by the properties of $V_{g}$. On the other hand, the first inequality in~(\ref{app_AgenH2}) is achievable by using symmetric states and jump operators (see supplementary material~\cite{supplement} for details) 
\beq
A_{\omega}(\rho^{\rm sym},\{L_{a,\omega}^{\rm sym}\}) = \sum_{k}p_{k}\mathcal{N}_{k}c_{k}(L_{a,\omega}^{\rm sym})F(\sigma^{\rm sym}_{j,k},B_{j,k}^{\mathrm{sym},\omega}). \label{app_Asymgen}
\eeq

}

As shown in Eqs.~(\ref{Asym}) and (\ref{app_Asymgen}), symmetric states and jump operators achieve the upper bound of the enhancement of $A_\omega$. It should be noted that the existence of symmetric states and jump operators requires condition $\dim(\mathcal{K}_{j}^{k})=1$ for one $j$, which we denote as $j_{\rm sym}$; note that $\sigma^{\rm sym}_{j,k}=B_{j,k}^{\mathrm{sym},\omega}=0$ for $j\neq j_{\rm sym}$ is satisfied for symmetric states and jump operators. This condition $\dim(\mathcal{K}_{j_{\rm sym}}^{k})=1$ is not necessarily satisfied for arbitrary $G$ and $V_g$. Therefore, this condition $\dim(\mathcal{K}_{j_{\rm sym}}^{k})=1$ can be viewed as a design principle of the Hamiltonian and jump operators to achieve the maximum enhancement of $A_{\omega}$. Note that the examples that we discuss in the main text satisfy this condition. 
We also note that when $\dim(\mathcal{H}_{j}^{k})=1$, $\rho_{k}^{\rm sym}$ is unique (if it exists) and can be written as $\rho^{\rm sym}_{k}=|\psi_{k}^{\rm sym}\rangle\langle \psi_{k}^{\rm sym}|$, where $|\psi_{k}^{\rm sym}\rangle$ is defined by $V_{g}|\psi_{k}^{\rm sym}\rangle = |\psi_{k}^{\rm sym}\rangle$ for all $g$.

\begin{figure}[t]
    \centering
    \includegraphics[width=0.95\linewidth]{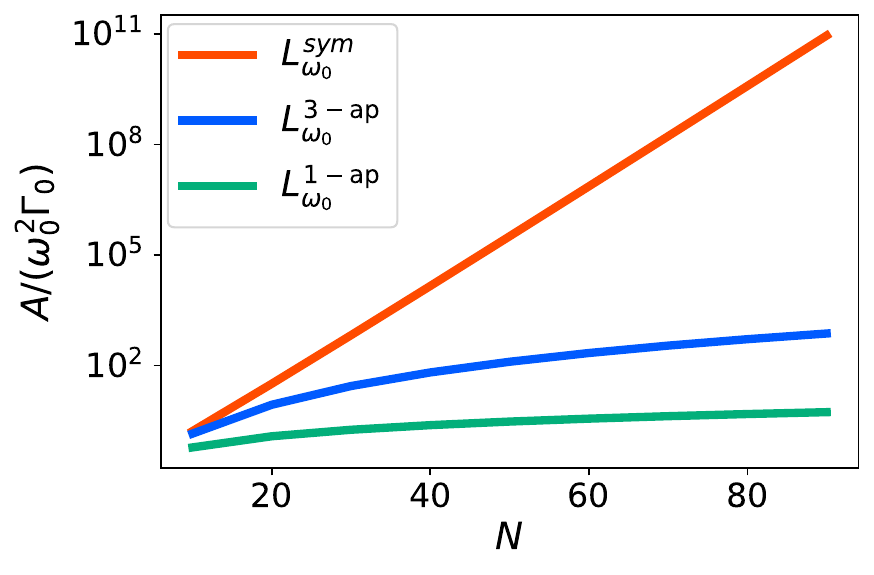}
    \caption{\revision{Scaling of $A$ for different jump operators when the density matrix is given by $\rho^{\rm sym}_{\rm th}$. The green curve is calculated by using $L^{1\text{-ap}}_{\pm\omega_{0}}$, the blue curve is calculated by using $L^{3\text{-ap}}_{\pm\omega_{0}}$, and the red curve is calculated by using $L_{\pm\omega_{0}}^{\rm sym}$. See also the supplementary material~\cite{supplement} for further details of the scaling of $A$. The parameters are $\omega_{0}=0.7,\beta=5$.} }
    \label{sup_fig_Athermal}
\end{figure}

\textbf{Appendix C: Scaling of $A$}

Because the density matrix during a heat engine cycle is generically given by a mixed state, here we consider the model discussed in Example 1 and show an additional plot that demonstrates the scaling of $A$ for $\rho^{\rm sym}_{\rm th}=\sum_{k}p_{k}^{\rm th}\rho_{k}^{\rm sym}$, where $p_{k}^{\rm th}=e^{-k\beta\omega_{0}}/\sum_{k=0}^{N}e^{-\beta k\omega_{0}}$ is the thermal occupation probability of the $k$-th energy eigenstate. Here, we consider the following master equation
\beq
\partial_{t}\rho = -i[H,\rho ]  + \gamma_{\downarrow} \mathcal{D}[L_{\omega_{0}}] \rho +\gamma_{\uparrow} \mathcal{D}[L_{\omega_{0}}^{\dagger}] \rho , \label{ME_app}
\eeq
where $L_{\omega_{0}}=\{L^{1\text{-ap}}_{\omega_{0}},L^{3\text{-ap}}_{\omega_{0}}, L^{\rm sym}_{\omega_{0}}\}$, $L_{\omega_{0}}^{\dagger}=L_{-\omega_{0}}$, $\gamma_{\downarrow}=\Gamma_{0}/(1+e^{-\beta\omega_{0}})$ and $\gamma_{\uparrow}=\Gamma_{0}/(1+e^{\beta\omega_{0}})$ satisfy the detailed balance condition $\gamma_{\downarrow}/\gamma_{\uparrow}=e^{\beta\omega_{0}}$. Note that $\rho_{\rm th}^{\rm sym}$ is the steady-state of Eq.~(\ref{ME_app}) when the initial state is prepared in the symmetric Dicke subspace (e.g., $\rho^{\rm sym}(0)=\sum_{k}p_{k}(0)\rho_{k}^{\rm sym}$), because $L^{1\text{-ap}}_{\pm\omega_{0}}$, $L^{3\text{-ap}}_{\pm\omega_{0}}$, and $L_{\pm\omega_{0}}^{\rm sym}$ satisfy the strong symmetry condition.  In Fig.~\ref{sup_fig_Athermal}, we plot $A=\sum_{\omega=\pm\omega_{0}}\omega^{2}A_{\omega}(\rho^{\rm sym}_{\rm th},L_{\omega})$, where its analytical expression, including the scaling of $A=O(N)$ for $L^{1\text{-ap}}_{\pm\omega_{0}}$ and $A=O(N^{3})$ for $L^{3\text{-ap}}_{\pm\omega_{0}}$ is obtained in the supplementary material~\cite{supplement}. From Fig.~\ref{sup_fig_Athermal}, we also find that $A$ scales exponentially for $L^{\rm sym}_{\pm\omega_{0}}$.

%\bibliography{ref}

\appendix
\widetext

\section{Details on the setup and main results}

\subsection{Symmetry condition}
Here, we summarize the symmetry condition on the jump operators introduced in the main text and show their relation. The master equation we consider is given by
\beq
\mathcal{L}\rho = -i[H,\rho] + \sum_{a,\omega}\gamma_{a,\omega} \Bigl( L_{a,\omega}\rho L^{\dagger}_{a,\omega}-\frac{1}{2}\{ L^{\dagger}_{a,\omega}L_{a,\omega} ,\rho \}  \Bigr), \label{sup_ME}
\eeq
and the jump operators are given by 
\beq
L_{a,\omega}=\sum_{\omega=\epsilon_{k}-\epsilon_{l}}\Pi_{l}S_{a}\Pi_{k}, \label{sup_jump}
\eeq
satisfying the relation $[\Pi_{k},L_{a,\omega}^{\dagger}L_{a,\omega}]=0$\revision{, where $\Pi_{k}$ is the projection to the $k$-th energy eigenspace $\mathcal{S}_{k}$.} 
We further assume the detailed balance condition $\gamma_{a,-\omega}=\gamma_{a,\omega}e^{-\beta \omega}$ to make Eq.~({\ref{sup_ME}) thermodynamically consistent, where $\beta$ is the inverse temperature of the heat bath. 

We assume that the Hamiltonian is invariant under the symmetry group 
\beq
[H,V_{g}]=0 \text{ for all } g, \label{HVcond}
\eeq
\revision{where $V_{g}$ is a unitary representation of the group. Note that from Eq.~(\ref{HVcond}), it follows that
\beq
[\Pi_{k},V_{g}]=0 \text{ for all } g, \label{sup_PikV}
\eeq
because for a given $k$-th energy eigenstate $|\psi_{k}\rangle$, the state $|\psi_{k}'\rangle = V_{g}|\psi_{k}\rangle$ is also a $k$-th energy eigenstate.
In addition, we impose the weak symmetry condition~(\ref{sup_Lweak}) on the jump operators. As we show below, the condition Eq.~(\ref{sup_LLsym}) follows from Eqs.~(\ref{HVcond}) and (\ref{sup_Lweak}), and we shall mainly use the property of Eq.~(\ref{sup_LLsym}) when deriving the main results.}   

Let us now summarize the conditions used in the main text and in the supplemental material as:
\begin{enumerate}
\item Symmetric jump operator condition
\beq
V_{g}L_{a,\omega}=L_{a,\omega}V^{\dagger}_{g} = L_{a,\omega} \ \text{ for all } g , \label{sup_LsymV1}
\eeq
\item Strong symmetry~\cite{Buca12}
\beq
[V_{g},L_{a,\omega}]=0  \ \text{ for all } g . \label{sup_Lstrong}
\eeq
\item Weak symmetry~\cite{Buca12}
\beq
\mathcal{L}(V_{g}xV^{\dagger}_{g})=V_{g}\mathcal{L}(x)V^{\dagger}_{g}\ \text{ for any operator } x, \text{ and for all } g . \label{sup_Lweak}
\eeq
\item \revision{Symmetry condition
\beq
\Bigl[\sum_{a}\gamma_{a,\omega}L_{a,\omega}^{\dagger}L_{a,\omega},V_{g}\Bigr]=0 \  \text{ for all } \omega\neq 0, \text{ and for all } g . \label{sup_LLsym}
\eeq
}
%By multiplying $\omega^{2}$ on both hands side of Eq.~(\ref{sup_LLsym}) and taking the summation over $\omega$, we also obtain the following condition:
%\beq
%\Bigl[ \sum_{a,\omega}\omega^{2}\gamma_{a,\omega}L_{a,\omega}^{\dagger}L_{a,\omega},V_{g}\Bigr]=0 \  \text{ for all } g . \label{sup_LLsym2}
%\eeq
\end{enumerate}
By assuming Eq.~(\ref{HVcond}), these conditions satisfy Eq.~(\ref{sup_LsymV1}) $\Rightarrow$ Eq.~(\ref{sup_Lstrong}) $\Rightarrow$ Eq.~(\ref{sup_Lweak}) $\Rightarrow$ Eq.~(\ref{sup_LLsym}).

Note that Eq.~(\ref{sup_LsymV1}) $\Rightarrow$ Eq.~(\ref{sup_Lstrong}) is straightforward. To show Eq.~(\ref{sup_Lstrong}) $\Rightarrow$ Eq.~(\ref{sup_Lweak}), we use the condition on the Hamiltonian $[H,V_{g}]=0$. 
\revision{Finally, we show Eq.~(\ref{sup_Lweak}) $\Rightarrow$ Eq.~(\ref{sup_LLsym}). To this end, we introduce the adjoint of the Lindblad super operator $\mathcal{L}^{\star}$ by the relation $\langle A,\mathcal{L}(B) \rangle=\langle \mathcal{L}^{\star}(A),B\rangle$, where $\langle A,B\rangle=\Tr[A^{\dagger}B]$ is the Hilbert-Schmidt scalar product. The explicit form of $\mathcal{L}^{\star}$ is given by 
\beq
\mathcal{L}^{\star}(A) = i[H,A] + \sum_{a,\omega}\gamma_{a,\omega}\Bigl( L_{a,\omega}^{\dagger}AL_{a,\omega} - \frac{1}{2}\{ L_{a,\omega}^{\dagger}L_{a,\omega},A\}\Bigr). \label{sup_Ladjoint}
\eeq
Using the weak symmetry condition~(\ref{sup_Lweak}), we find that 
\beq
\langle A, \mathcal{L}(V_{g}BV_{g}^{\dagger})\rangle = \langle A, V_{g}\mathcal{L}(B)V_{g}^{\dagger} \rangle. \label{sup_Ladj1}
\eeq
The left-hand side of Eq.~(\ref{sup_Ladj1}) can be further calculated as
\beq
\langle A, \mathcal{L}(V_{g}BV_{g}^{\dagger})\rangle = \langle \mathcal{L}^{\star}(A), V_{g}BV_{g}^{\dagger}\rangle = \langle  V_{g}^{\dagger}\mathcal{L}^{\star}(A)V_{g}, B \rangle . \label{sup_Ladj2}
\eeq
Similarly, the right-hand side of Eq.~(\ref{sup_Ladj1}) can be further calculated as
\beq
\langle A, V_{g}\mathcal{L}(B)V_{g}^{\dagger} \rangle =\langle V_{g}^{\dagger}AV_{g}, \mathcal{L}(B) \rangle = \langle \mathcal{L}^{\star}(V_{g}^{\dagger}AV_{g}), B \rangle. \label{sup_Ladj3}
\eeq
By combining Eqs.~(\ref{sup_Ladj1}), (\ref{sup_Ladj2}), and (\ref{sup_Ladj3}), we obtain the corresponding symmetry condition on $\mathcal{L}^{\star}$ as
\beq
\mathcal{L}^{\star}(V_{g}^{\dagger}xV_{g})=V_{g}^{\dagger}\mathcal{L}^{\star}(x)V_{g}\ \text{ for any operator } x, \text{ and for all } g . \label{sup_Lweak_ad}
\eeq
Next, we use Eq.~(\ref{sup_Lweak_ad}) and obtain
\beq
\Pi_{k}\mathcal{L}^{\star}(\Pi_{l})\Pi_{k} = \Pi_{k}\mathcal{L}^{\star}(V_{g}^{\dagger}\Pi_{l}V_{g})\Pi_{k} = \Pi_{k} V_{g}^{\dagger} \mathcal{L}^{\star}(\Pi_{l})V_{g}\Pi_{k} . \label{sup_Ladj4}
\eeq
By using Eqs.~(\ref{sup_jump}) and (\ref{sup_Ladjoint}), the left-hand side of Eq.~(\ref{sup_Ladj4}) reads
\beq
\Pi_{k}\mathcal{L}^{\star}(\Pi_{l})\Pi_{k}=\sum_{a}\gamma_{a,\epsilon_{k}-\epsilon_{l}}L_{a,\epsilon_{k}-\epsilon_{l}}^{\dagger}L_{a,\epsilon_{k}-\epsilon_{l}} -\delta_{k,l} \Pi_{k}\sum_{a,\omega} \gamma_{a,\omega} L_{a,\omega}^{\dagger}L_{a,\omega}\Pi_{k} \label{sup_Ladj5},
\eeq
and the right-hand side of Eq.~(\ref{sup_Ladj4}) reads
\beq
\Pi_{k} V_{g}^{\dagger} \mathcal{L}^{\star}(\Pi_{l})V_{g}\Pi_{k} = V_{g}^{\dagger} \sum_{a}\gamma_{a,\epsilon_{k}-\epsilon_{l}}L_{a,\epsilon_{k}-\epsilon_{l}}^{\dagger}L_{a,\epsilon_{k}-\epsilon_{l}} V_{g}-\delta_{k,l} V_{g}^{\dagger}\Pi_{k}\sum_{a,\omega} \gamma_{a,\omega} L_{a,\omega}^{\dagger}L_{a,\omega}\Pi_{k}V_{g}. \label{sup_Ladj6}
\eeq
By combining Eqs.~(\ref{sup_Ladj4}), (\ref{sup_Ladj5}), and (\ref{sup_Ladj6}) and by denoting $\omega=\epsilon_{k}-\epsilon_{l}$, we obtain
\beq
\sum_{a}\gamma_{a,\omega}L_{a,\omega}^{\dagger}L_{a,\omega} = 
V_{g}^{\dagger} \sum_{a}\gamma_{a,\omega}L_{a,\omega}^{\dagger}L_{a,\omega} V_{g},
\eeq
for $\omega\neq 0$, which shows Eq.~(\ref{sup_LLsym}). 
}

\subsection{Irreducible decomposition}

\revision{In this subsection, we give give the irreducible decomposition of $V_{g}$ discussed in the main text. First, from Eq.~(\ref{sup_PikV}), the unitary operator $V_{g}$ can be decomposed as
\beq
V_{g} = \bigoplus_{k} V_{g}^{k},
\eeq
where each $V_{g}^{k}$ acts on the $k$-th eigenspace $\mathcal{S}_{k}$. Now, each $V_{g}^{k}$ can be decomposed into irreducible representations as
\beq
V^{k}_{g}=\bigoplus_{j} I_{\mathcal{H}_{j}^{k}}\otimes \mathcal{V}^{k}_{j}(g), \label{sup_irreps}
\eeq
where $\mathcal{V}_{j}^{k}(g)$ is an irreducible representation of $G$ acting on the subspace $\mathcal{K}_{j}^{k}$, $I_{\mathcal{H}_{j}^{k}}$ is the identity matrix acting on the subspace $\mathcal{H}_{j}^{k}$, and $j$ labels irreducible representations. 
Here, the $k$-th energy eigenspace $\mathcal{S}_{k}$ is decomposed as $\mathcal{S}_{k}=\bigoplus_{j} \mathcal{H}_{j}^{k}\otimes \mathcal{K}_{j}^{k}$, which means that the subspace $\mathcal{H}_{j}^{k}$ is invariant under the operation of $V^{k}_{g}$, and $\mathcal{K}_{j}^{k}$ is the only subspace in which $V_{g}^{k}$ nontrivially acts on. 

If $\dim(\mathcal{H}_{j}^{k})\geq 2$, the chosen group $G$ does not completely specify the symmetry of the Hamiltonian, in general. It is possible to construct unitary operators that nontrivially act on the subspace $\mathcal{H}_{j}^{k}$, which are related to unspecified groups $G'$ (see, for example, Sec.~\ref{sec:Perm}).  

More generally, for given Hamiltonian $H$, group $G$, and its unitary representation $V_{g}$ satisfying $[H,V_{g}]=0$, we can appropriately choose a group $G'$ and $\{\tilde{V}_{(g,g')}\}_{(g,g')\in G\times G'}$ such that $[\tilde{V}_{(g,g')},H]=0$ and the irreducible decomposition $\tilde{V}_{(g,g')}=\bigoplus_{j}I_{\tilde{\mathcal{H}}_{j}^{k}}\otimes \tilde{\mathcal{V}}_{j}^{k}(g,g')$ with $\dim(\tilde{\mathcal{H}}_{j}^{k}) =1$ for any $j,k$ is satisfied for $G\times G'$ and its unitary representation $\tilde{V}_{(g,g')}$, where $\tilde{\mathcal{V}}_{j}^{k}(g,g')$ acts on the subspace $\tilde{\mathcal{K}}_{j}^{k}$, and the $k$-th energy eigenspace is decomposed as $\mathcal{S}_{k}=\bigoplus_{j}\tilde{\mathcal{H}}_{j}^{k}\otimes \tilde{\mathcal{K}}_{j}^{k}$. To show this claim, start from the decomposition~(\ref{sup_irreps}). For each $\mathcal{H}_{j}^{k}$, we choose a group $G'_{(k,j)}$ and its unitary representation $\{U_{g'(k,j)}\}_{g'(k,j)\in G'(k,j)}$, such that $U_{g'(k,j)}$ is an irreducible representation acting on $\mathcal{H}_{j}^{k}$; for example, $SU(\dim(\mathcal{H}_{j}^{k}))$ and its natural representation satisfies this property. We then take $G'=\times_{(k,j)}G'_{(k,j)}$ and $\tilde{V}_{(g,g')}=\oplus_{k}\oplus_{j}U_{g'_{(k,j)}}\otimes \mathcal{V}_{j}^{k}(g)$. We now note that because $U_{g'_{(k,j)}}$ and $\mathcal{V}_{j}^{k}(g)$ are both irreducible, $U_{g'(k,j)}\otimes \mathcal{V}_{j}^{k}(g)$ is also an irreducible representation of $G\times G'_{(k,j)}$ acting on $\mathcal{H}_{j}^{k}\otimes \mathcal{K}_{j}^{k}$~\cite{etingof2011introductionrepresentationtheory}. %In doing so, consider a operator $D$ that satisfies $[D,U_{g'(k,j)}\otimes \mathcal{V}_{j}^{k}(g)]=0$ for any $(g,g'(k,j))\in G\times G'_{(k,j)}$. Then, by taking $g'(k,j)$ as the identity element, we have $[D,I_{\mathcal{H}_{j}^{k}}\otimes \mathcal{V}_{j}^{k}(g)]=0$ for any $g$. Because $\mathcal{V}_{j}^{k}(g)$ is an irreducible representation acting on $\mathcal{K}_{j}^{k}$, $D$ must take the form $D=D'\otimes I_{\mathcal{K}_{j}^{k}}$, where $D'$ is an operator acting on $\mathcal{H}_{j}^{k}$. Similarly, by taking $g$ as the identity element, we have $[D'\otimes I_{\mathcal{K}_{j}^{k}},U_{g'(k,j)}\otimes I_{\mathcal{H}_{j}^{k}}]=0$ for any $g'(k,j)$. Because $U_{g'(k,j)}$ is an irreducible representation acting on $\mathcal{H}_{j}^{k}$, we have $D'=c I_{\mathcal{H}_{j}^{k}}$, where $c$ is a constant. We thus conclude that $D$ must be $D=cI_{\mathcal{H}_{j}^{k}\otimes \mathcal{K}_{j}^{k}}$, which completes the proof that $U_{g'(k,j)}\otimes \mathcal{V}_{j}^{k}(g)$ is an irreducible representation acting on $\mathcal{H}_{j}^{k}\otimes \mathcal{K}_{j}^{k}$. 
Therefore, we have $\tilde{\mathcal{K}}_{j}^{k}=\mathcal{H}_{j}^{k}\otimes \mathcal{K}_{j}^{k}$ and thus $\dim(\tilde{\mathcal{H}}_{j}^{k})=1$. In summary, we can always choose an appropriate group and its unitary representation that satisfies $\dim(\mathcal{H}_{j}^{k})=1$. 
}

\revision{
Because the subspace $\mathcal{H}_{j}^{k}$ is invariant under the operation of $V_{g}$, we cannot constrain the form of the quantum states and jump operators within this invariant subspace by imposing conditions based on $V_{g}$. Specifically, there exists freedom in choosing the operators $\sigma_{j,k}$ and $B_{j,k}^{\omega}$ that act on the subspace $\mathcal{H}_{j}^{k}$ as shown in Eqs.~(\ref{sup_qstate2}) and (\ref{sup_bjump2}). By recalling that our main aim is to classify quantum states and jump operators based on $V_{g}$, we shall first consider the case of choosing an appropriate $G$ and $V_{g}$ and assume $\dim(\mathcal{H}_{j}^{k})=1$. We also discuss the general case $\dim(\mathcal{H}_{j}^{k})\geq 2$ in Sec.~\ref{sec:Ageneral}. 
}

\subsection{List of the local and symmetric states and jump operators}
\revision{For clarity, let us list the definitions of local and symmetric states and jump operators introduced in the main text:
\begin{itemize}
\item Local states $\rho_{k}^{\rm loc}$, defined by
\beq
\frac{1}{|G|}\sum_{g\in G}V_{g}\rho_{k}^{\rm loc}V_{g}^{\dagger}=\frac{1}{\mathcal{N}_{k}} \Pi_{k} . \label{sup_localBlock}
\eeq
We also denote $\rho^{\rm loc}=\sum_{k}p_{k}\rho_{k}^{\rm loc}$, where $p_{k}$ is the occupation probability of the $k$-th energy eigenstates. 

\item Local energy eigenbasis for $\mathcal{S}_{k}$, defined by
\beq
\mathcal{B}^{\rm loc}_{k}=\{ |\psi_{k}^{\rm loc}(\alpha)\rangle\}_{\alpha=1}^{\mathcal{N}_{k}},
\eeq
where each element of $\mathcal{B}^{\rm loc}_{k}$ satisfies Eq.~(\ref{sup_localBlock}).

\item Symmetric states
\beq
    V_{g}\rho_{k}^{\rm sym}= \rho_{k}^{\rm sym}V_{g}^{\dagger} = \rho_{k}^{\rm sym} \ \text{ for all } g. \label{sup_defSymState}
    \eeq
    %This definition means that $\rho_{k}^{\rm sym}$ is an operator acting on the subspace $\mathcal{H}_{j_{\rm sym}}^{k}\otimes \mathcal{K}_{j_{\rm sym}}^{k}$, where $\dim(I^{k}_{\mathcal{K}_{j_{\rm sym}}})=1$. We denote the label $j$ that satisfies this condition as $j_{\rm sym}$. 

\item Local jump operators
\beq
\Bigl[(L^{\rm loc}_{a,\omega})^{\dagger}L^{\rm loc}_{a,\omega},|\psi_{k}^{\rm loc}(\alpha)\rangle\langle \psi_{k}^{\rm loc}(\alpha)|\Bigr]=0, \text{ for all } \alpha. \label{sup_Lloccond1}
\eeq

\item Symmetric jump operators
   \beq
    V_{g}L^{\rm sym}_{\omega}=L^{\rm sym}_{\omega}V^{\dagger}_{g} = L^{\rm sym}_{\omega} \ \text{ for all } g . \label{sup_LsymV}
    \eeq 

\end{itemize}

We note that the existence of symmetric states requires the condition $\dim(\mathcal{K}_{j}^{k})=1$ for one $j$. Otherwise, since $\mathcal{V}_{j}^{k}(g)$ nontrivially acts on the subspace $\mathcal{K}_{j}^{k}$, the condition~(\ref{sup_defSymState}) cannot be satisfied. In the following, we denote the label $j$ satisfying $\dim(\mathcal{K}_{j}^{k})=1$ as $j_{\rm sym}$. Then, $\rho_{k}^{\rm sym}$ can be expressed as Eq.~(\ref{appendix:symstate}). As noted in the main text, this condition is not necessarily satisfied for general $G$ and $V_{g}$.  

We note that local states and symmetric states are different in general. Suppose $\rho_{k}$ satisfies both Eqs.~(\ref{sup_localBlock}) and (\ref{sup_defSymState}). If we substitute Eq.~(\ref{sup_defSymState}) into Eq.~(\ref{sup_localBlock}), we find that $\rho_{k}=\Pi_{k}/\mathcal{N}_{k}$; however, this contradicts with Eq.~(\ref{sup_defSymState}). Similarly, the conditions for local jump operators~(\ref{sup_Lloccond1}) and symmetric jump operators~(\ref{sup_LsymV}) cannot be satisfied simultaneously. Suppose $L_{\omega}$ satisfies Eqs.~(\ref{sup_Lloccond1}) and (\ref{sup_LsymV}). From Eq.~(\ref{sup_Lloccond1}), the jump operator takes the form $L_{\omega}^{\dagger}L_{\omega}=\sum_{k,\alpha}l^{\omega}_{k,\alpha}|\psi_{k}^{\rm loc}(\alpha)\rangle\langle \psi_{k}^{\rm loc}(\alpha)|$, where $l_{k,\alpha}^{\omega}$ is some coefficient. Then, by using Eq.~(\ref{sup_LsymV}), we find that $L_{\omega}^{\dagger}L_{\omega}=|G|^{-1}\sum_{g}V_{g}\sum_{k,\alpha}l^{\omega}_{k,\alpha}|\psi_{k}^{\rm loc}(\alpha)\rangle\langle \psi_{k}^{\rm loc}(\alpha)|V_{g}^{\dagger}=\sum_{k}(\sum_{\alpha}l_{k,\alpha}^{\omega}) \Pi_{k}/\mathcal{N}_{k}$, but this contradicts with Eq.~(\ref{sup_LsymV}).

%We also note that $\dim(\mathcal{H}_{j}^{k})=1$ is a necessary condition for the existence of $\mathcal{B}_{k}^{\rm loc}$. Suppose that $\dim(\mathcal{H}_{j}^{k})\geq 2$, and write the basis set as $|a\rangle\otimes |b\rangle$, where $|a\rangle\in \mathcal{H}_{j}^{k}$ and $|b\rangle\in\mathcal{K}_{j}^{k}$. Then, $|\psi_{k}^{\rm loc}\rangle=\sum_{a,b}$

%Note that the local basis set $\mathcal{B}_{k}^{\rm loc}$ can be constructed as follows. 
%Let us assume $\dim(\mathcal{H}_{j}^{k})=1$ and introduce a basis set $\{|j,k\rangle\otimes|i_{j}\rangle\}$ for $\mathcal{S}_{k}$, where $\{|i_{j}\rangle\}_{i_{j}=1}^{d_{j}^{k}}$ is a basis set for $\mathcal{K}_{j}^{k}$, and $d_{j}^{k}=\dim(\mathcal{K}_{j}^{k})$. We find that 
%\beq
%\frac{1}{|G|}\sum_{g}V_{g}|j,k,i\rangle\langle j,k,i|V^{\dagger}_{g}=|j,k\rangle\langle j,k|\otimes \frac{1_{\mathcal{K}_{j}^{k}}}{d_{j}^{k}}.
%\eeq 
%Therefore, $|\psi_{k}^{\rm loc}(\alpha)\rangle$ can be constructed by taking an appropriate superposition of $|j,k,i\rangle$ by choosing the weights of each $j$ as $\sqrt{d_{j}^{k}/\mathcal{N}_{k}}$. One example is $|\psi^{\rm loc}_{k}(\alpha)\rangle=\sum_{j}\sqrt{d_{j}^{k}/\mathcal{N}_{k}}|j,k,i\rangle$.  
}

\section{Derivation of the main results}

\revision{In this section, we show the main results Eqs.~(7) to (10) presented in the main text. As discussed in the main text, we assume $\dim(\mathcal{H}_{j}^{k})=1$ from Sec.~\ref{sec:Aloc} to Sec.~\ref{sec:Asym}. We show the case of $\dim(\mathcal{H}_{j}^{k})\geq 2$ in Sec.~\ref{sec:Ageneral}, and discuss how the main results get modified by this generalization. 

Because our main focus is on the scaling of the activity-like quantity $A=\sum_{\omega\neq 0}\omega^{2}A_{\omega}$, we discuss the scaling of $A_{\omega}$ for $\omega\neq 0$ in the following. 

We also note that, as it is apparent from the derivation given below, the main results Eqs.~(7) to (10) do not rely on the detailed balance condition $\gamma_{a,-\omega}=\gamma_{a,\omega}e^{-\beta \omega}$; however, this condition is typically assumed in the (quantum) thermodynamics settings and is also assumed in the derivation of the current-dissipation trade-off relation $2J^{2}/\dot{\sigma} \leq A$~\cite{Tajima21}. Therefore, we shall assume the detailed balance condition throughout the paper. 
}

\subsection{\label{sec:Aloc}$A_{\omega}$ for local states}

\revision{We show Eq.~(7) in the main text:
\beq
A_{\omega}(\rho^{\rm loc},\{L_{a,\omega}\})=\sum_{k}p_{k}c_{k}(L_{a,\omega}).  \label{sup_Alocstate} 
\eeq
We first note that 
\beq
A_{\omega}(\rho^{\rm loc},\{L_{a,\omega}\}) = \sum_{k}p_{k}\Tr\Bigl[\sum_{a}\gamma_{a,\omega}L_{a,\omega}^{\dagger}L_{a,\omega}\rho_{k}^{\rm loc} \Bigr] =\sum_{k}p_{k}\Tr\Bigl[V_{g}^{\dagger}\sum_{a}\gamma_{a,\omega}L_{a,\omega}^{\dagger}L_{a,\omega}V_{g}\rho^{\rm loc}_{k} \Bigr]  ,
\eeq
where we use Eq.~(\ref{sup_LLsym}) and obtain the second equality. Using the cyclic property of the trace and taking summation over $g$, we obtain
\beqa
A_{\omega}(\rho^{\rm loc},\{L_{a,\omega}\}) &=& \sum_{k}p_{k}\Tr\Bigl[\sum_{a}\gamma_{a,\omega}L_{a,\omega}^{\dagger}L_{a,\omega} \frac{1}{|G|}\sum_{g\in G}V_{g}\rho^{\rm loc}_{k} V_{g}^{\dagger}\Bigr]\nonumber \\
&=& \sum_{k}p_{k}\frac{1}{\mathcal{N}_{k}}\Tr\Bigl[\sum_{a}\gamma_{a,\omega}L_{a,\omega}^{\dagger}L_{a,\omega} \Pi_{k} \Bigr] = \sum_{k}p_{k}c_{k}(L_{a,\omega}),
\eeqa
where we use Eq.~(\ref{sup_localBlock}) and obtain the second line. Therefore, Eq.~(\ref{sup_Alocstate}) is obtained.
}

\subsection{$A_{\omega}$ for local jump operators}

\revision{We next show Eq.~(8) in the main text:
\beq
A_{\omega}(\rho,\{L_{a,\omega}^{\rm loc}\}) = \sum_{k}p_{k}c_{k}(L_{a,\omega}^{\rm loc}). \label{sup_Alocjump}
\eeq

To show Eq.~(\ref{sup_Alocjump}), we first show that the local jump operators $\{L^{\rm loc}_{a,\omega}\}$ satisfy the following relation
\beq
\sum_{a}\gamma_{a,\omega}(L^{\rm loc}_{a,\omega})^{\dagger}L^{\rm loc}_{a,\omega} =\sum_{k}c_{k}(L_{a,\omega}^{\rm loc}) \Pi_{k} \label{LdagLdecom}
\eeq
for $\omega\neq 0$, because by assuming Eq.~(\ref{LdagLdecom}), we have
\beq
A_{\omega}(\rho,\{L_{a,\omega}^{\rm loc}\}) = \sum_{k}p_{k} \Tr\Bigl[\sum_{a}\gamma_{a,\omega}(L^{\rm loc}_{a,\omega})^{\dagger}L^{\rm loc}_{a,\omega}\rho_{k} \Bigr] = \sum_{k}p_{k}c_{k}(L_{a,\omega}^{\rm loc}) \Tr[\rho_{k}\Pi_{k}] = \sum_{k}p_{k}c_{k}(L_{a,\omega}^{\rm loc}) ,
\eeq
and Eq.~(\ref{sup_Alocjump}) can be obtained. Therefore, in what follows, we show Eq.~(\ref{LdagLdecom}). 

To show Eq.~(\ref{LdagLdecom}), we take summation over $a$ on both hand sides of the definition of the local jump operators~(\ref{sup_Lloccond1}):
\beq
\Bigl[\sum_{a}(L^{\rm loc}_{a,\omega})^{\dagger}L^{\rm loc}_{a,\omega},|\psi_{k}^{\rm loc}(\alpha)\rangle\langle \psi_{k}^{\rm loc}(\alpha)|\Bigr]=0, \text{ for all } \alpha. \label{sup_Lloccond1}
\eeq
From the above relation, we have
\beq
\sum_{a}\gamma_{a,\omega}(L^{\rm loc}_{a,\omega})^{\dagger}L^{\rm loc}_{a,\omega} = \sum_{k} \sum_{\alpha} A^{\alpha}_{k} |\psi_{k}^{\rm loc}(\alpha)\rangle\langle \psi_{k}^{\rm loc}(\alpha)|,
\eeq
where $A^{\alpha}_{k}$ is some coefficient. Using the commutation relation for jump operators Eq.~(\ref{sup_LLsym}), we have
\beq
\sum_{a}\gamma_{a,\omega}(L^{\rm loc}_{a,\omega})^{\dagger}L^{\rm loc}_{a,\omega} = \frac{1}{|G|}\sum_{g\in G}V_{g} \Bigl( \sum_{a}\gamma_{a,\omega}(L^{\rm loc}_{a,\omega})^{\dagger}L^{\rm loc}_{a,\omega}\Bigr) V_{g}^{\dagger} = \frac{1}{|G|}\sum_{g\in G} V_{g}\sum_{k}\sum_{\alpha} A^{\alpha}_{k} |\psi_{k}^{\rm loc}(\alpha)\rangle\langle \psi_{k}^{\rm loc}(\alpha)|V_{g}^{\dagger}.
\eeq
We then use the relation~(\ref{sup_localBlock}) for local states 
\beq
\frac{1}{|G|}\sum_{g\in G}V_{g}|\psi_{k}^{\rm loc}(\alpha)\rangle\langle \psi_{k}^{\rm loc}(\alpha)|V_{g}^{\dagger}=\frac{1}{\mathcal{N}_{k}} \Pi_{k}, \label{sup_locdensity}
\eeq
and obtain
\beq
\sum_{a}\gamma_{a,\omega}(L^{\rm loc}_{a,\omega})^{\dagger}L^{\rm loc}_{a,\omega} = \sum_{k}\sum_{\alpha}A_{k}^{\alpha} \frac{1}{\mathcal{N}_{k}} \Pi_{k}. \label{sup:Localproof1}
\eeq
We further note that 
\beq
c_{k}(L_{a,\omega}^{\rm loc}) =\frac{1}{\mathcal{N}_{k}}\sum_{a}\gamma_{a,\omega}\Tr[ \Pi_{k}(L^{\rm loc}_{a,\omega})^{\dagger}L^{\rm loc}_{a,\omega}\Pi_{k}]=\sum_{\alpha}A_{k}^{\alpha}\frac{1}{\mathcal{N}_{k}^{2}}\Tr[\Pi_{k}]= \frac{1}{\mathcal{N}_{k}}\sum_{\alpha}A_{k}^{\alpha}, \label{sup:Localproof2}
\eeq
and by substituting Eq.~(\ref{sup:Localproof2}) into Eq.~(\ref{sup:Localproof1}) shows Eq.~(\ref{LdagLdecom}). 
}

\subsection{General upper bound of $A_{\omega}$}

\revision{We now show a general upper bound on $A_{\omega}$ (Eq. (9) in the main text):
\beq
A_{\omega}(\rho,L_{a,\omega})  \leq \sum_{k}p_{k}\mathcal{N}_{k}c_{k}(L_{a,\omega}), \label{sup_supAgen}
\eeq
for any states $\rho$ and any jump operators $ L_{a,\omega}$. 

To show Eq.~(\ref{sup_supAgen}), we first note that 
\beqa
A_{\omega}(\rho,\{L_{a,\omega}\}) &=& \sum_{k}p_{k}\Tr\Bigl[\sum_{a}\gamma_{a,\omega}L_{a,\omega}^{\dagger}L_{a,\omega}\rho_{k} \Bigr] \nonumber \\
&=& \sum_{k}p_{k}\Tr\Bigl[V_{g}^{\dagger}\sum_{a}\gamma_{a,\omega}L_{a,\omega}^{\dagger}L_{a,\omega}V_{g}\rho_{k} \Bigr] \nonumber \\
&=& \sum_{k}p_{k}\Tr\Bigl[\sum_{a}\gamma_{a,\omega}L_{a,\omega}^{\dagger}L_{a,\omega}\frac{1}{|G|}\sum_{g\in G}V_{g}\rho_{k}V_{g}^{\dagger} \Bigr] \label{sup_ALrho} ,
\eeqa
where the second equality is obtained by using the commutation relation~(\ref{sup_LLsym}). Because the density matrix $\frac{1}{|G|}\sum_{g\in G}V_{g}\rho_{k}V_{g}^{\dagger}$ commutes with $V_{g}$, we can express it using the block-diagonal form based on the irreducible decomposition~(\ref{sup_irreps}), which now reads
\beq
\frac{1}{|G|}\sum_{g\in G}V_{g}\rho_{k}V_{g}^{\dagger} =\frac{1}{\sum_{j}q_{j,k}} \bigoplus_{j} q_{j,k} |j,k\rangle\langle j,k|\otimes \frac{I_{\mathcal{K}_{j}^{k}}}{d_{j}^{k}}, \label{sup_qstate}
\eeq
where we use $\dim(\mathcal{H}_{j}^{k})=1$ and denote the basis in $\mathcal{H}_{j}^{k}$ as $|j,k\rangle$, $q_{j,k}$ is a coefficient, and $d_{j}^{k}=\dim(\mathcal{K}_{j}^{k})$.

Similarly, the jump operators satisfy the commutation relation~(\ref{sup_LLsym}), and can be block-diagonalized as
\beq
\sum_{a}\gamma_{a,\omega}L_{a,\omega}^{\dagger}L_{a,\omega} = \bigoplus_{k,j}b_{j,k}^{\omega}|j,k\rangle\langle j,k|\otimes \frac{I_{\mathcal{K}_{j}^{k}}}{d_{j}^{k}}, \label{sup_bjump}
\eeq
where $b_{j,k}^{\omega}$ is a coefficient. 

Using Eqs.~(\ref{sup_qstate}) and (\ref{sup_bjump}), $A_{\omega}$ can be expressed as
\beq
A_{\omega}(\rho,\{L_{a,\omega}\}) = \sum_{k}p_{k}\frac{\sum_{j}(d_{j}^{k})^{-1}q_{j,k}b_{j,k}^{\omega}}{\sum_{j}q_{j,k}}\leq \sum_{k}p_{k} \frac{\sum_{j}q_{j,k}b_{j,k}^{\omega}}{\sum_{j}q_{j,k}}\leq \sum_{k}p_{k}\sum_{j}b_{j,k}^{\omega}, \label{sup_Abdjk}
\eeq
where the second inequality follows from 
\beq
(d_{j}^{k})^{-1}\leq 1 \text{ for all }j,\label{sup_eqcond}
\eeq
and the last inequality follows from 
\beq
\frac{q_{j,k}}{\sum_{j}q_{j,k}}\leq 1 \text{ for all }j. \label{sup_eqcond2}
\eeq
Finally, we note that 
\beq
c_{k}(L_{a,\omega})=\frac{1}{\mathcal{N}_{k}}\Tr\Bigl[\sum_{a}\gamma_{a,\omega}L_{a,\omega}^{\dagger}L_{a,\omega} \Pi_{k} \Bigr] = \frac{1}{\mathcal{N}_{k}}\sum_{j}b_{j,k}^{\omega}. \label{sup_ckgen}
\eeq
By combining Eqs.~(\ref{sup_Abdjk}) and (\ref{sup_ckgen}), we obtain the main result~(\ref{sup_supAgen}).

}

\subsection{\label{sec:Asym}$A_{\omega}$ for symmetric state and jump operator}
\revision{We now show that by using the symmetric states $\rho^{\rm sym}$ and jump operators $\{L_{a,\omega}^{\rm sym}\}$, $A_{\omega}$ is given by Eq.~(10) in the main text:
\beq
\mathcal{A}_{\omega}(\rho^{\rm sym},L^{\rm sym}_{a,\omega}) =  \sum_{k}p_{k} \mathcal{N}_{k} c_{k}(L_{a,\omega}^{\rm sym}). \label{supAsym}
\eeq 

To show Eq.~(\ref{supAsym}), we start from the conditions~(\ref{sup_defSymState}) and (\ref{sup_LsymV}), and express the symmetric state $\rho^{\rm sym}_{k}$ and jump operator $L_{a,\omega}^{\rm sym}$ as
\beqa
\rho^{\rm sym}_{k} &=& |j_{\rm sym},k\rangle\langle j_{\rm sym},k|\otimes I_{\mathcal{K}_{j_{\rm sym}}^{k}} , \label{appendix:symstate} \\
\sum_{a}\gamma_{a,\omega}(L_{a,\omega}^{\rm sym})^{\dagger}L_{a,\omega}^{\rm sym} &=& \bigoplus_{k}b_{j_{\rm sym},k}^{\omega} |j_{\rm sym},k\rangle\langle j_{\rm sym},k| \otimes I_{\mathcal{K}_{j_{\rm sym}}^{k}},
\eeqa
where $d_{j_{\rm sym}}^{k}=\dim(\mathcal{K}_{j_{\rm sym}}^{k})=1$. A direct calculation reads to
\beq
A_{\omega}(\rho^{\rm sym},L_{a,\omega}^{\rm sym}) = \sum_{k}p_{k}b_{j_{\rm sym},k}^{\omega}. \label{sup_Asymgen1}
\eeq
Combining Eq.~(\ref{sup_Asymgen1}) with 
\beq
c_{k}(L_{a,\omega}^{\rm sym})=\frac{1}{\mathcal{N}_{k}}b_{j_{\rm sym},k}^{\omega}
\eeq
shows the result~(\ref{supAsym}). We further note that because $q_{j_{\rm sym},k}=1$, $q_{j,k}=0$ for $j\neq j_{\rm sym}$, and $d_{j_{\rm sym}}^{k}=1$, the equality conditions in Eqs.~(\ref{sup_eqcond}) and (\ref{sup_eqcond2}) are satisfied, and thus the upper bound of~(\ref{sup_supAgen}) is achieved.

}

\subsection{\label{sec:Ageneral}The case when $\dim(\mathcal{H}_{j}^{k})\geq 2$}
\revision{Finally, we discuss how the results presented in the main text is modified when invariant subspace under the action of $V_{g}^{k}$ exists, i.e., $\dim(\mathcal{H}_{j}^{k})\geq 2$. The main results Eqs.~(9) and (10) are now generalized to Eqs.~(\ref{sup_AgenH1}) and (\ref{sup_Asymgen}), respectively. 

We start by using Eq.~(\ref{sup_ALrho}), since it is still valid when $\dim(\mathcal{H}_{j}^{k})\geq 2$. However, Eqs.~(\ref{sup_qstate}) and (\ref{sup_bjump}) get modified as follows:
\beqa
\frac{1}{|G|}\sum_{g\in G}V_{g}\rho_{k}V_{g}^{\dagger} &=& \frac{1}{\sum_{j}\Tr[\sigma_{j,k}]}\bigoplus_{j} \sigma_{j,k} \otimes \frac{I_{\mathcal{K}_{j}^{k}}}{d_{j}^{k}},  \label{sup_qstate2} \\
\sum_{a}\gamma_{a,\omega}L_{a,\omega}^{\dagger}L_{a,\omega} &=& \bigoplus_{k,j}B_{j,k}^{\omega} \otimes \frac{I_{\mathcal{K}_{j}^{k}}}{d_{j}^{k}}, \label{sup_bjump2}
\eeqa
where $\sigma_{j,k}$ and $B_{j,k}^{\omega}$ are operators acting on the subspace $\mathcal{H}_{j}^{k}$. Using the expressions~(\ref{sup_qstate2}) and (\ref{sup_bjump2}), we obtain
\beq
A_{\omega}(\rho,\{L_{a,\omega}\}) = \sum_{k}p_{k} \frac{\sum_{j}(d_{j}^{k})^{-1}\Tr[\sigma_{j,k}B_{j,k}^{\omega}]}{\sum_{j}\Tr[\sigma_{j,k}]}\leq \sum_{k}p_{k} \frac{\sum_{j}\Tr[\sigma_{j,k}B_{j,k}^{\omega}]}{\sum_{j}\Tr[\sigma_{j,k}]}, \label{sup_Abdjk2}
\eeq
where we use the inequality~(\ref{sup_eqcond}) and obtain the second inequality. We further note that
\beq
c_{k}(L_{a,\omega})=\frac{1}{\mathcal{N}_{k}}\sum_{j}\Tr[B_{j,k}^{\omega}]. \label{sup_Abdjk3}
\eeq
Combining Eqs.~(\ref{sup_Abdjk2}) and (\ref{sup_Abdjk3}) gives an upper bound on the average jump rate
\beqa
A_{\omega}(\rho,\{L_{a,\omega}\}) &\leq& \sum_{k}p_{k}\mathcal{N}_{k}c_{k}(L_{a,\omega}) \frac{\sum_{j}\Tr[\sigma_{j,k}B_{j,k}^{\omega}]}{\sum_{j}\Tr[\sigma_{j,k}]\sum_{j}\Tr[B_{j,k}^{\omega}]} \label{sup_AgenH1} \\ 
&\leq &\sum_{k}p_{k}\mathcal{N}_{k}c_{k}(L_{a,\omega}) , \label{sup_AgenH2}
\eeqa
which generalizes the result Eq.~(9) to the case of $\dim(\mathcal{H}_{j}^{k})\geq 2$. Note that the last inequality~(\ref{sup_AgenH2}) is obtained by noting that
\beq
\frac{\sum_{j}\Tr[\sigma_{j,k}B_{j,k}^{\omega}]}{\sum_{j}\Tr[\sigma_{j,k}]\sum_{j}\Tr[B_{j,k}^{\omega}]} \leq 1, \label{sup_AgenH_eq}
\eeq
and thus the equality condition in Eq.~(\ref{sup_AgenH2}) requires specific conditions on the form of $\sigma_{j,k}$ and $B_{j,k}^{\omega}$. However, the subspace $\mathcal{H}_{j}^{k}$ is invariant under $V_{g}^{k}$, which means that one cannot characterize the equality condition in Eq.~(\ref{sup_AgenH2}) by using the properties of $V_{g}^{k}$. Therefore, the upper bound generically depends on the specific form of $\sigma_{j,k}$ and $B_{j,k}^{\omega}$, shown by the tighter inequality~(\ref{sup_AgenH1}). 

Next, we show that the equality condition of~(\ref{sup_AgenH1}) can be achieved by using the symmetric states and jump operators. From the conditions~(\ref{sup_defSymState}) and (\ref{sup_LsymV}), the symmetric state $\rho^{\rm sym}_{k}$ and jump operators $L_{a,\omega}^{\rm sym}$ take the form
\beqa
\rho^{\rm sym}_{k} &=& \frac{1}{\Tr[\sigma_{j_{\rm sym},k}]}\sigma_{j_{\rm sym},k}\otimes I_{\mathcal{K}_{j_{\rm sym}}^{k}} \\
\sum_{a}\gamma_{a,\omega}(L_{a,\omega}^{\rm sym})^{\dagger}L_{a,\omega}^{\rm sym} &=& \bigoplus_{k}B_{j_{\rm sym},k}^{\omega}\otimes I_{\mathcal{K}_{j_{\rm sym}}^{k}},
\eeqa
where $d_{j_{\rm sym}}^{k}=\dim(\mathcal{K}_{j_{\rm sym}}^{k})=1$. Note that depending on the freedom of the choice of $\sigma_{j_{\rm sym},k}$, there could be multiple symmetric states. A direct calculation reads to
\beq
A_{\omega}(\rho^{\rm sym},L_{a,\omega}^{\rm sym}) = \sum_{k}p_{k}\frac{\Tr[\sigma_{j_{\rm sym},k}B_{j_{\rm sym},k}^{\omega}]}{\Tr[\sigma_{j_{\rm sym},k}]}= \sum_{k}p_{k}\mathcal{N}_{k}c_{k}(L_{a,\omega}^{\rm sym})\frac{\Tr[\sigma_{j_{\rm sym},k}B_{j_{\rm sym},k}^{\omega}]}{\Tr[\sigma_{j_{\rm sym},k}]\Tr[B_{j_{\rm sym},k}^{\omega}]}, \label{sup_Asymgen}
\eeq
by noting that 
\beq
c_{k}(L_{a,\omega}^{\rm sym})=\frac{1}{\mathcal{N}_{k}}\Tr[B_{j_{\rm sym},k}^{\omega}].
\eeq
Therefore, the upper bound in~(\ref{sup_AgenH1}) is achieved by noting that $\sigma_{j,k}=B_{j,k}^{\omega}=0$ for $j\neq j_{\rm sym}$, and the equality condition in~(\ref{sup_eqcond}) is achieved since $d_{j_{\rm sym}}^{k}=1$. Therefore, the enhancement of $A_{\omega}$ for symmetric states and jump operators presented in Eq.~(10) is generalized to Eq.~(\ref{sup_Asymgen}) when $\dim(\mathcal{H}_{j}^{k})\geq 2$, as it depends on specific forms of $\sigma_{j_{\rm sym},k}$ and $B_{j_{\rm sym},k}^{\omega}$. 
}

\section{Details on the permutaiton-invariant two-level systems}
In this section, we give further details on the identical two-level systems. 
\subsection{Permutation symmetry and block-diagonal structures}
The Hamiltonian of the system is given by
\beq
H=\hbar\omega\sum_{i}^{N}\sigma^{z}_{i},
\eeq
which is permutation-invariant. \revision{By denoting the permutation group as $S_{N}$ and a unitary representation of the group as $V_{g}$ $(g\in S_{N})$, the Hamiltonian satisfies $[H,V_{g}]=0$. Then, the $k$-th eigenspace is decomposed as
\beq
\mathcal{S}_{k}=\bigoplus_{j=|j_{\rm sym}-k|}^{j_{\rm sym}} \mathcal{H}_{j}^{k}\otimes \mathcal{K}_{j}^{k},
\eeq
where $j$ labels the irreducible representation of $S_{N}$, $j_{\rm sym}=N/2$, and $k=0,\cdots,N$. The projection operator $\Pi_{k}$, which satisfies $[\Pi_{k},V_{g}]=[\Pi_{k},V_{g}^{k}]=0$ with $V_{g}=\bigoplus_{k}V_{g}^{k}$, can be decomposed as
\beq
\Pi_{k} = \bigoplus_{j=|j_{\rm sym}-k|}^{j_{\rm sym}} |j,k\rangle \langle j,k| \otimes I_{\mathcal{K}_{j}^{k}},
\eeq
where the basis $|j,k\rangle \in \mathcal{H}_{j}^{k}$ is the eigenstate of $J^{2}=J_{x}^{2}+J_{y}^{2}+J_{z}^{2}$ and $J_{z}$, with $J_{\alpha}=\sum_{i}\sigma^{\alpha}_{i}$ for $\alpha=x,y,z$, i.e., $J^{2}|j,k\rangle=j(j+1)|j,k\rangle$ and $J_{z}|j,k\rangle=(k-N/2)|j,k\rangle$. 
From the above expression, this model satisfies the condition
\beq
\dim(\mathcal{H}_{j}^{k})=1. \label{sup_Hjk1}
\eeq
We also note that $\dim(\mathcal{K}_{j}^{k})={}_{N}C_{N/2-j}- {}_{N}C_{N/2-j-1}$ and $\dim(\mathcal{S}_{k})=\mathcal{N}_{k}={}_{N}C_{k}=\sum_{j=|N/2-k|}^{N/2}\dim(\mathcal{K}_{j}^{k})$. Moreover, $\dim(\mathcal{K}_{j_{\rm sym}}^{k})=1$, and the basis 
\beq
|\psi_{k}^{\rm sym}\rangle = |j_{\rm sym},k\rangle = \frac{1}{\sqrt{{}_{N}C_{k}}}\sum_{g\in S_{N}} V_{g}^{k} |e\rangle^{\otimes k}\otimes |g\rangle^{\otimes N-k}, \label{sup_symDicke}
\eeq
is known as the symmetric Dicke states, satisfying $V_{g}|\psi^{\rm sym}_{k}\rangle = V_{g}^{k}|\psi^{\rm sym}_{k}\rangle = |\psi^{\rm sym}_{k}\rangle$ for all $g\in S_{N}$. Therefore, $\rho_{k}^{\rm sym}=|\psi_{k}^{\rm sym}\rangle\langle\psi_{k}^{\rm sym}|$ satisfies the condition~(\ref{sup_defSymState}). 
}

\subsection{symmetric jump operator}
The symmetric jump operator given in the main text reads
\beq
L^{\rm sym}_{\omega_{0}}= \sum_{m=0}^{\lceil N/2\rceil-1}L^{(m)}_{\omega_{0}}   , \label{sup_SR-symL}
\eeq
and its approximation by taking the first $2n+1$-body terms reads
\beq
L^{2n+1\text{-ap}}_{\omega_{0}}=\sum_{m=0}^{n}L^{(m)}_{\omega_{0}}.
\eeq
Here, the $L^{(m)}$ is defined as
\beq
L^{(m)}_{\omega_{0}}=\sum_{(i_{1}<\cdots<i_{m+1})\neq (k_{1}<\cdots <k_{m})}\sigma_{i_{1}}^{-}\cdots \sigma_{i_{m+1}}^{-}\sigma_{k_{1}}^{+}\cdots \sigma_{k_{m}}^{+}.
\eeq

To calculate the average jump rate, we use the following relation
\beq
L^{(m)}_{\omega_{0}}|\psi^{\rm sym}_{k}\rangle = \sqrt{k(N-k+1)}[1+f(N,k,m)]|\psi^{\rm sym}_{k-1}\rangle , \label{Lm_factor}
\eeq
where
\beq
f(N,k,m)= \frac{(N-k)\cdots (N-k+1-m)}{(m+1)!}\frac{(k-1)\cdots (k-m)}{m!}, \label{sup_prefactor}
\eeq
and $f(N,k,m=0)=0$. The prefactor in front of Eq.~(\ref{Lm_factor}) can be understood as follows. If we fix $\sigma_{i_{1}}^{-}$, other $\sigma^{-}$'s acting on the Dicke state have the multiplicity $(k-1)\cdots (k-m)$, and $\sigma^{+}$'s acting on the Dicke state have the multiplicity $(N-k)\cdots (N-k+1-m)$. Finally, the factor $k(N-k+1)$ arises from the Klein-Gordon coefficient for SU(2) operators: $J_{-}|\psi_{k}^{\rm sym}\rangle=\sqrt{k(N-k+1)}|\psi_{k-1}^{\rm sym}\rangle$, with $J_{-}=\sum_{i}\sigma_{i}^{-}$. Using Eq.~(\ref{Lm_factor}), we find that 
\beq
A_{\omega_{0}}(\rho^{\rm sym},L^{2n+1\text{-ap}}_{\omega_{0}})=c_{k}(L^{2n+1\text{-ap}}_{\omega_{0}})\sum_{m=0}^{n} {}_{N-k+1} C_{m+1} \cdot  {}_{k-1}C_{k-1-m}, \label{enhapp}
\eeq
and 
\beq
c_{k}(L^{2n+1\text{-ap}}_{\omega_{0}})=\frac{k}{N-k+1}\sum_{m=0}^{n} {}_{N-k+1} C_{m+1} \cdot  {}_{k-1}C_{k-1-m} \gamma_{\downarrow}.
\eeq

\revision{
Finally, we take $n=\lceil N/2\rceil-1$ and show the relation 
\beq
A_{\omega_{0}}(\rho^{\rm sym},L^{\rm sym}_{\omega_{0}}) = c_{k}(L^{\rm sym}_{\omega_{0}}){}_{N}C_{k}, \label{sup_enh_sym}
\eeq
and
\beq
c_{k}(L^{\rm sym}_{\omega_{0}}) = \frac{k}{N-k+1} {}_{N}C_{k} \gamma_{\downarrow}.
\eeq
If $k\geq \lceil N/2\rceil$, we find that $\sum_{m=0}^{\lceil N/2\rceil-1} {}_{N-k+1} C_{m+1} \cdot  {}_{k-1}C_{k-1-m}=\sum_{m=0}^{N-k}{}_{N-k+1} C_{N-k-m} \cdot  {}_{k-1}C_{m}$. By using Vandermonde's identity $\sum_{m=0}^{a}{}_{b} C_{m} \cdot  {}_{c}C_{a-m}= {}_{b+c}C_{a}$ for $a=N-k, b=k-1,c=N-k+1$, we obtain Eq.~(\ref{sup_enh_sym}). Next, we consider the case $k\leq \lceil N/2\rceil$. In this case, we find that $\sum_{m=0}^{\lceil N/2\rceil-1} {}_{N-k+1} C_{m+1} \cdot  {}_{k-1}C_{k-1-m}=\sum_{m=0}^{k-1}{}_{N-k+1} C_{m+1} \cdot  {}_{k-1}C_{k-1-m}=\sum_{j=1}^{k}{}_{N-k+1} C_{j} \cdot  {}_{k-1}C_{k-j}$, where the last equality is obtained by changing the variable to $j=m+1$. We then extend the summation to the case of $j=0$, since ${}_{k-1}C_{k}=0$. Finally, we use Vandermonde's identity and obtain Eq.~(\ref{sup_enh_sym}).  
}

\subsection{local jump operators}
There are many choices for the local jump operators. In particular, for each 
\beq
L^{2n+1\text{-ap}}_{a,\omega_{0}}=\sum_{i}\sigma_{i}^{-} + \sum_{(i_{1}<i_{2})\neq l_{1}}\sigma_{i_{1}}^{-}\sigma_{i_{2}}^{-}\sigma_{l_{1}}^{+} + \cdots + \sum_{(i_{1}<\cdots<i_{n+1})\neq (l_{1}<\cdots<l_{n})} \sigma_{i_{1}}^{-}\cdots \sigma_{i_{n+1}}^{-}\sigma_{l_{1}}^{+}\cdots \sigma_{l_{n}}^{+} , \label{sup_ap}
\eeq
we can construct a local jump operator with the same normalization factor $c_{k}$. The explicit form of the local jump operators that contain up to $2n+1$-body jump operators read 
\beq
\{ L^{2n+1\text{-loc}}_{a,\omega_{0}} \} = \{ \{ \sigma_{i}^{-} \}, \{ \sigma_{i_{1}}^{-}\sigma_{i_{2}}^{-}\sigma_{l_{1}}^{+} \},\cdots, \{ \sigma_{i_{1}}^{-}\cdots \sigma_{i_{n+1}}^{-}\sigma_{l_{1}}^{+}\cdots \sigma_{l_{n}}^{+} \} \} . \label{sup_loc}
\eeq
Each element of Eq.~(\ref{sup_loc}) satisfies the condition Eq.~(\ref{sup_localBlock}). The master equation reads
\beq
\partial_{t}\rho = -i[H,\rho] + \gamma_{\downarrow} \Bigl( \sum_{i}\mathcal{D}[\sigma_{i}]  + \sum_{(i_{1}<i_{2})\neq l_{1}} \mathcal{D}[\sigma_{i_{1}}^{-}\sigma_{i_{2}}^{-}\sigma_{l_{1}}^{+}] + \cdots \sum_{(i_{1}<\cdots<i_{n+1})\neq (l_{1}<\cdots<l_{n})} \mathcal{D}[\sigma_{i_{1}}^{-}\cdots \sigma_{i_{n+1}}^{-}\sigma_{l_{1}}^{+}\cdots \sigma_{l_{n}}^{+}] \Bigr)\rho, \label{Local2nME}
\eeq
where $\mathcal{D}[L]\rho=L\rho L^{\dagger}-(1/2)\{L^{\dagger}L,\rho\}$ is the dissipator. Note that Eq.~(\ref{Local2nME}) satisfies the weak symmetry~\cite{Buca12}. An argument similar to that below Eq.~(\ref{sup_prefactor}) leads to   
\beq
A_{\omega_{0}}(\rho,\{ L^{2n+1\text{-loc}}_{a,\omega_{0}}  \}) = \sum_{k}p_{k} c_{k}(\{ L^{2n+1\text{-loc}}_{a,\omega_{0}}  \}),
\eeq
and $c_{k}(\{ L^{2n+1\text{-loc}}_{a,\omega_{0}}  \})=c_{k}(L^{2n+1\text{-ap}}_{\omega_{0}})$. Note that the jump operators act individually on the system in the case of $\{ L^{2n+1\text{-loc}}_{a,\omega_{0}}\}$, whereas they act collectively in the case of $L^{2n+1\text{-ap}}_{\omega_{0}}$. They share same number of operators involved in their expression, and thus the Hilbert-Schmidt norm of the jump operators $c_{k}(\{ L^{2n+1\text{-loc}}_{a,\omega_{0}}  \})$ and $c_{k}(L^{2n+1\text{-ap}}_{\omega_{0}})$ take the same value.

\subsection{Scaling of $A$}
\revision{We now discuss the scaling behavior of $A$. Let us first assume that the explicit form of the master equation reads
\beq
\partial_{t}\rho = -i[H,\rho ]  + \gamma_{\downarrow} \mathcal{D}[L_{\omega_{0}}] \rho +\gamma_{\uparrow} \mathcal{D}[L_{\omega_{0}}^{\dagger}] \rho , \label{ME_sup2}
\eeq
with $L_{\omega_{0}}=\{L^{1\text{-ap}}_{\omega_{0}},L^{3\text{-ap}}_{\omega_{0}}, L^{\rm sym}_{\omega_{0}}\}$, $\gamma_{\downarrow}=\Gamma_{0}/(1+e^{-\beta\omega_{0}})$ and $\gamma_{\uparrow}=\Gamma_{0}/(1+e^{\beta\omega_{0}})$ satisfy the detailed balance condition $\gamma_{\downarrow}/\gamma_{\uparrow}=e^{\beta\omega_{0}}$. These jump operators satisfy the strong symmetry condition, and therefore the steady-state depends on the initial state. If the initial state is prepared in the so-called symmetric Dicke subspace (e.g., $\rho(0)=\sum_{k,j}\rho_{k,j}|\psi_{k}^{\rm sym}\rangle\langle \psi_{j}^{\rm sym}|$), the steady-state is given by the thermal state within this subspace:
\beq
\rho^{\rm sym}_{\rm th}=\sum_{k}p_{k}^{\rm th}|\psi_{k}^{\rm sym}\rangle\langle \psi_{k}^{\rm sym}|, \label{sup_sym_th}
\eeq
where $p_{k}^{\rm th}=e^{-k\beta\omega}/Z$ and $Z=\sum_{k=0}^{N}e^{-k\beta\omega}$. 

The explicit form of $A=\sum_{\omega=\pm\omega_{0}}\omega^{2}A_{\omega}(\rho^{\rm sym},L_{\omega})$, with $\rho^{\rm sym}=\sum_{k}p_{k}|\psi_{k}^{\rm sym}\rangle\langle\psi_{k}^{\rm sym}|$ is given by
\beqa
A^{1\text{-ap}} &=& \omega_{0}^{2}\sum_{k}p_{k} [ \gamma_{\downarrow}k(N-k+1)+ \gamma_{\uparrow}(k+1)(N-k)], \label{sup_A1scaling} \\
A^{3\text{-ap}} &=&\omega_{0}^{2}\sum_{k}p_{k} \left( \gamma_{\downarrow}k(N-k+1)[1+(k-1)(N-k)/2]^{2}+\gamma_{\uparrow}(k+1)(N-k)[1+k(N-k-1)/2]^{2}\right), \label{sup_A3scaling} \\
A^{\rm sym} &=& \omega_{0}^{2}\sum_{k}p_{k}\Bigl[\gamma_{\downarrow} ({}_{N}C_{k})^{2}\frac{k}{N-k+1}+\gamma_{\uparrow}( {}_{N}C_{k+1})^{2}\frac{k+1}{N-k} \Bigr]. \label{sup_Asymscaling}
\eeqa
In Fig.~\ref{sup_fig_Athermal}, we plot Eqs.~(\ref{sup_A1scaling}), (\ref{sup_A3scaling}), and (\ref{sup_Asymscaling}) when the density matrix of the system is given by $\rho^{\rm sym}_{\rm th}$ [Eq.~(\ref{sup_sym_th})]. Note that for large $N$, the scaling behavior $A^{1\text{-ap}}=O(N)$ and $A^{3\text{-ap}}=O(N^{3})$ can be analyticaly obtained as well by noting that $\sum_{k}p_{k}^{\rm th}k^{n}=O(1)$ for $n=0,1,\cdots$. The red curve shows that $A^{\rm sym}$ scales exponentially.

Next, we assume that the jump operators are given by $\{\sigma_{i}^{-}\}$ and $\{\sigma_{i}^{+}\}$. The master equation reads
\beq
\partial_{t}\rho = -i[H,\rho ]  + \gamma_{\downarrow} \sum_{i=1}^{N}\mathcal{D}[\sigma_{i}^{-}] \rho +\gamma_{\uparrow} \sum_{i=1}^{N}\mathcal{D}[\sigma_{i}^{+}] \rho . \label{ME_sup3}
\eeq
In this case, the jump operator does not satisfy the strong symmetry condition, and the steady-state of Eq.~(\ref{ME_sup3}) is uniquely given by a thermal state
\beq
\rho_{\rm th}= \frac{e^{-\beta H}}{\Tr[e^{-\beta H}]} = \sum_{k} p_{k}^{\rm th}\Pi_{k}. \label{sup_rhoth}
\eeq
Because Eq.~(\ref{sup_rhoth}) satisfies the condition~(\ref{sup_localBlock}), we find that $A_{\omega}(\rho_{\rm th},\{L_{a,\omega}\})=\sum_{k}p_{k}^{\rm th}c_{k}(L_{a,\omega})$ for any jump operators. The explicit form of $A$ using jump operators $\{\sigma_{i}^{-}\}$ and $\{\sigma_{i}^{+}\}$ reads
\beq
A^{\rm loc} = \omega_{0}^{2}\sum_{k}p_{k}[\gamma_{\downarrow}k + \gamma_{\uparrow}(N-k)].
\eeq
}

\begin{figure}
    \centering
    \includegraphics[width=0.7\linewidth]{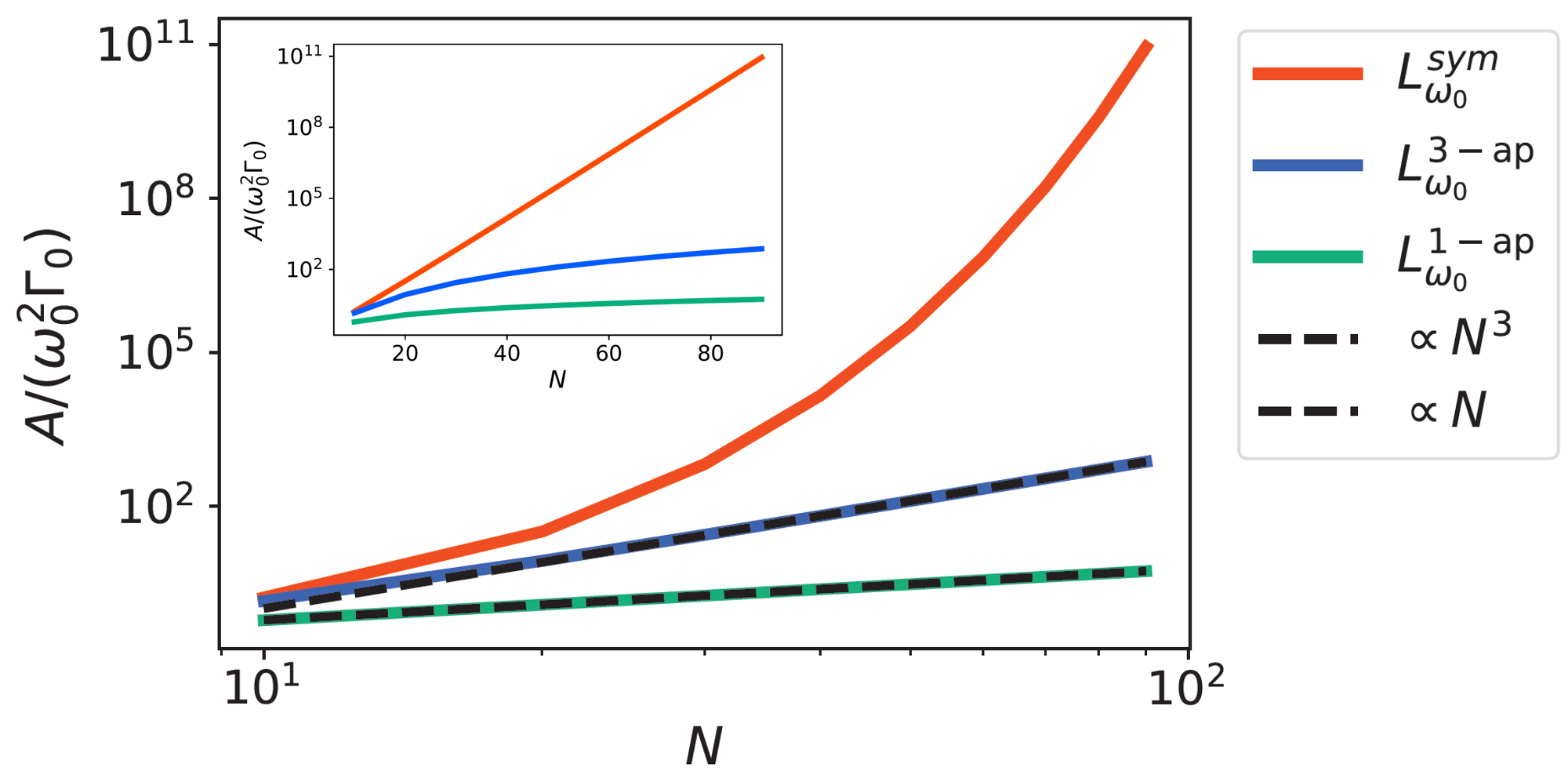}
    \caption{\revision{Scaling of $A$ for different jump operators when the density matrix is given by $\rho^{\rm sym}_{\rm th}$~(\ref{sup_sym_th}). The figure shows that $A$ scales $O(N)$ for $L^{1\text{-ap}}_{\pm\omega_{0}}$ and $O(N^{3})$ for $L^{3\text{-ap}}_{\pm\omega_{0}}$. We use a semi-log plot in the inset, which shows an exponential scaling of $A$ for the symmetric jump operator $L_{\pm\omega_{0}}^{\rm sym}$. The parameters are $\omega_{0}=0.7,\beta=5$.} }
    \label{sup_fig_Athermal}
\end{figure}

\section{Details on the quantum heat engine model}
In this section we provide details on the quantum heat engine model discussed in the main text. The time evolution of the system reads
\beq
\partial_{t}\rho = \mathcal{L}_{j}\rho = -i[H_{j},\rho ] + \sum_{a,\omega}\gamma_{a,\omega}^{j} \mathcal{D}[L_{a,\omega}] \rho , \label{ME_sup} 
\eeq
where $j=\{H,C\}$ labels whether the system is interacting with the hot bath ($H$) or the cold bath ($C$), and $H_{j}=(\omega_{j}/2) \sum_{i=1}^{N}\sigma_{i}^{z}$ is the Hamiltonian of the system. We assume the detailed balance condition $\gamma_{a,\omega}^{j}=e^{\beta_{j}\omega_{j}}\gamma_{a,-\omega}^{j}$, where $\beta_{j}$ is the inverse temperature of the heat bath. 

We consider a standard quantum Otto cycle, where the four-stroke protocol reads 
\begin{enumerate}
    \item Investing work via changing the frequency of the system $\omega_{C}\rightarrow \omega_{H}$. 
    \item Absorbing heat $Q_{H}\geq 0$ with the hot bath at inverse temperature $\beta_{H}$ for a time-duration $\tau_{H}$, using Eq.~(\ref{ME_sup}).
    \item Extracting work via changing the frequency of the system $\omega_{H} \rightarrow \omega_{C}$. 
    \item Releasing heat $Q_{C}\geq 0$ to the cold bath at inverse temperature $\beta_{C}$ for a time-duration $\tau_{C}$, using Eq.~(\ref{ME_sup}). 
\end{enumerate}

Using the first law of thermodynamics, the extracted work reads $W=Q_{H}-Q_{C}$. The performance of the heat engine is quantified by the heat-to-work conversion efficiency $\eta=W/Q_{H}$, and the output power $P=W/(\tau_{H}+\tau_{C})$. In steps 1 and 3, we assume that the system is detached from the heat bath. \revision{Suppose that we write the time-dependence of the Hamiltonian during step 1 as $H(t)=(\omega(t)/2)\sum_{i=1}^{N}\sigma_{i}^{z}$, where $\omega(t)$ denotes the time-dependence of changing the frequency of the Hamiltonian from $\omega(t_{\rm ini})=\omega_{C}$ to $\omega(t_{\rm fin})=\omega_{H}$. We note that the Hamiltonian of the system during step 1 (and also step 3) commutes with each other at different times, i.e., $[H(t),H(t')]=0$. In particular, the occupation probabilities $p_{k}$'s of the energy eigenstates are unchanged during steps 1 and 3, regardless of the speed of changing $\omega(t)$. Therefore, typical detrimental effects on the power and efficiency of heat engines that occur for fast and non-commuting Hamiltonian protocols ($[H(t),H(t')]\neq 0$) do not occur in the above protocol. From the above consideration, we assume that steps 1 and 3 can be completed instantaneously, and identify the total cycle time as $\tau_{H}+\tau_{C}$. See also the analysis of a sudden cycle~\cite{PhysRevB.100.085405} which corresponds to a heat engine protocol with steps 1 and 3 done instantaneously for both commuting and non-commuting Hamiltonians.} The above protocol realizes a heat engine cycle when $\beta_{H}\omega_{H} \geq \beta_{C}\omega_{C}$.

When the engine cycle becomes stationary, the density matrix at the beginning of step 1 satisfies the condition $\rho_{1}=e^{\mathcal{L}_{C}\tau_{C}}e^{\mathcal{L}_{H}\tau_{H}}\rho_{1}$. The work and heat read 
\beq
W=(\omega_{H}-\omega_{C})\sum_{i=1}^{N}\Tr[ \sigma_{i}^{z} (e^{\mathcal{L}_{H}\tau_{H}}-1)\rho_{1}]. \label{Work}
\eeq
Similarly, the heat reads
\beqa
Q_{H} &=& \omega_{H}\sum_{i=1}^{N}\Tr[ \sigma_{i}^{z} (e^{\mathcal{L}_{H}\tau_{H}}-1)\rho_{1}] \label{QH} \\
Q_{C} &=& -\omega_{C}\sum_{i=1}^{N}\Tr[ \sigma_{i}^{z} (e^{\mathcal{L}_{C}\tau_{C}}e^{\mathcal{L}_{H}\tau_{H}}-e^{\mathcal{L}_{H}\tau_{H}})\rho_{1}].
\eeqa
From Eqs.~(\ref{Work}) and (\ref{QH}), the efficiency reads 
\beq
\eta = \frac{W}{Q_{H}}= 1-\frac{\omega_{C}}{\omega_{H}},
\eeq
which is identical to the Otto efficiency. By choosing 
\beq
\omega_{C}=\frac{\beta_{H}}{\beta_{C}} \omega_{H} + \frac{b}{N}\omega_{H}, \label{omegaC}
\eeq
the difference between the Carnot efficiency $\eta_{\rm Car}=1-\beta_{H}/\beta_{C}$ and the efficiency reads 
\beq
\eta_{\rm Car}-\eta = \frac{b}{N} = O(1/N). \label{sup_Carnot}
\eeq

In the numerical simulation, we fix  $b=0.1, \omega_{H}=0.7, \beta_{H}=5, \beta_{C}=8$ and use Eq.~(\ref{omegaC}) to determine $\omega_{C}$. We also consider the form $\gamma_{a,\omega_{j}}^{j}=\gamma_{\downarrow}^{j}=\Gamma_{0}/(1+e^{-\beta_{j}\omega_{j}})$ and $\gamma_{a,-\omega_{j}}^{j}=\gamma_{\uparrow}^{j}=\Gamma_{0}/(1+e^{\beta_{j}\omega_{j}})$, and set $\Gamma_{0}=0.2$ in the numerical simulation. We numerically optimize the output power for $N=1$ by varying $\tau_{H}$ and $\tau_{C}$, and find that $\tau_{H}^{N=1}=0.168$, $\tau_{C}^{N=1}=0.180$. For $N>2$, we choose $\tau_{H}=\tau_{H}^{N=1}N^{-3}$ and $\tau_{C}=\tau_{C}^{N=1}N^{-3}$ for $L^{1\text{-ap}}_{\omega_{0}}$ and $L^{3\text{-ap}}_{\omega_{0}}$. We also choose $\tau_{H}=\tau_{H}^{N=1}N^{-6}$ and $\tau_{C}=\tau_{C}^{N=1}N^{-6}$ for $L_{\omega_{0}}^{\rm sym}$. Using these parameters, we obtain the numerical plot in the main text by considering three different jump operators $L^{1\text{-ap}}_{\omega_{0}}$, $L^{3\text{-ap}}_{\omega_{0}}$, and $L^{\rm sym}_{\omega_{0}}$. Here, we assume that the initial state is prepared by the symmetric Dicke state. Then, the time-evolution of the system is confined in the subspace spanned by the symmetric Dicke states by noting that all jump operators satisfy the strong symmetry condition~\cite{Buca12}. In Fig.~\ref{sup_fig1}, we show additional plots of the output power by using the local jump operators. 

\begin{figure}
    \centering
    \includegraphics[width=0.5\linewidth]{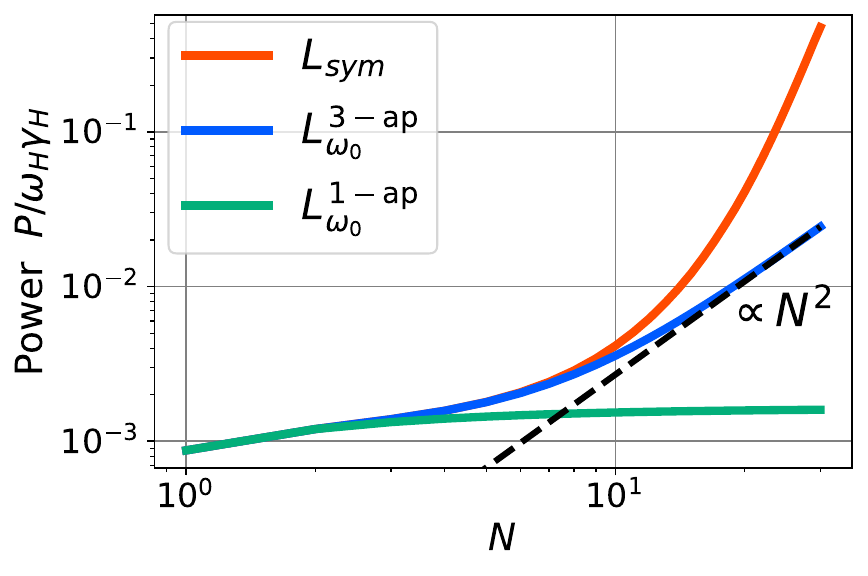}
    \caption{Plot of the power of the permutation invariant $N$ two-level system heat engine using the log-log plot. Here, the black dashed line is added to show the $N^2$ scaling of the power when we use $L^{3\text{-ap}}_{\omega_{0}}$. 
    }
    \label{sup_fig2}
\end{figure}

\begin{figure}
    \centering
    \includegraphics[width=0.95\linewidth]{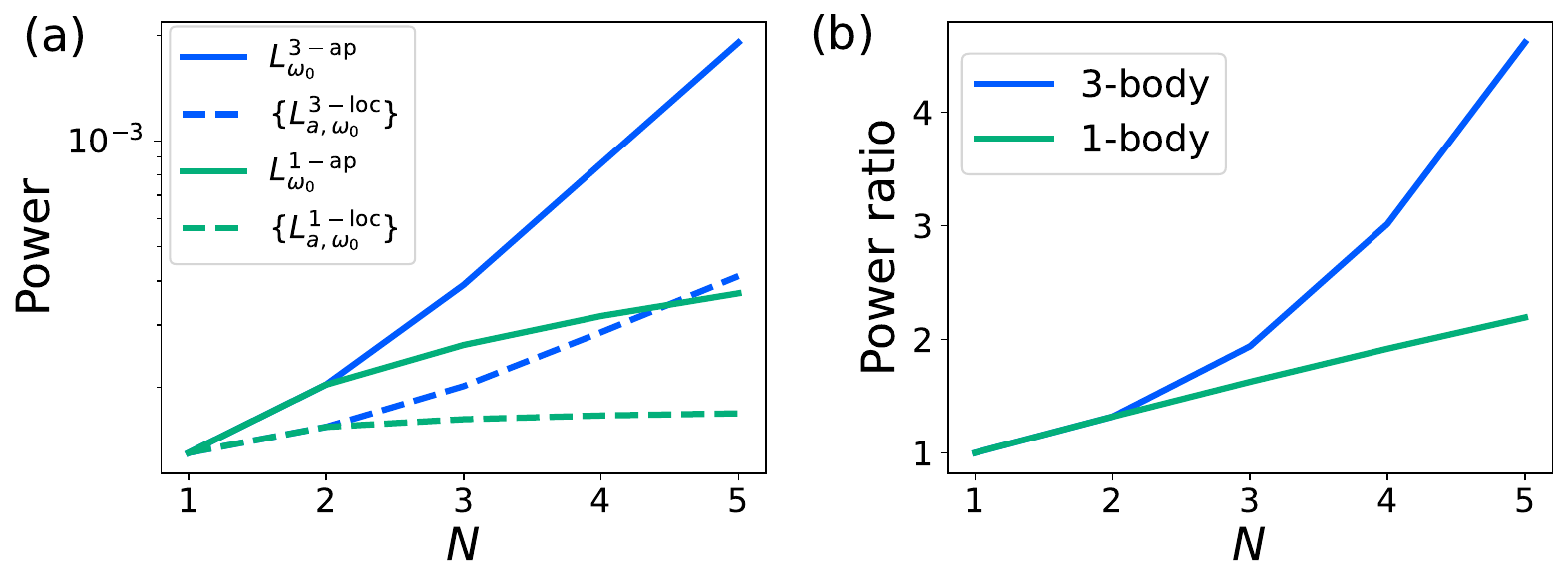}
    \caption{(a) Plot of the output power of heat engines for different jump operators $L^{3\text{-ap}}_{\omega_{0}}$, $\{L^{3\text{-loc}}_{a,\omega_{0}}\}$, $L^{1\text{-ap}}_{\omega_{0}}$, $\{L^{1\text{-loc}}_{a,\omega_{0}}\}$. (b) Plot of the ratio of the power between jump operators $L^{2n+1-\text{ap}}_{\omega_{0}}$ and $\{L^{2n+1-\text{loc}}_{a,\omega_{0}} \}$ for 1-body ($n=0$) and 3-body ($n=1$) terms. The blue curve is given by $P^{3-\text{ap}}/P^{3-\text{loc}}$ and the green curve is given by $P^{1-\text{ap}}/P^{1-\text{loc}}$. The parameters are $b=0.1,\omega_{H}=0.7,\beta_{H}=0.5, \beta_{C}=0.8, \Gamma_{0}=0.2$.}
    \label{sup_fig1}
\end{figure}

The explicit form of the master equation reads
\beq
\partial_{t}\rho = -i[H_{j},\rho ]  + \gamma_{\downarrow}^{j} \mathcal{D}[L_{\omega_{0}}] \rho +\gamma_{\uparrow}^{j} \mathcal{D}[L_{\omega_{0}}^{\dagger}] \rho , \label{ME_sup1}
\eeq
with $L_{\omega_{0}}=\{L^{1\text{-ap}}_{\omega_{0}},L^{3\text{-ap}}_{\omega_{0}}, L^{\rm sym}_{\omega_{0}}\}$.

The analytical expression of $\bar{A}$ using $L^{1\text{-ap}}_{\omega_{0}}$ reads 
\beq
\bar{A}^{1\text{-ap}}=\omega_{0}^{2}\sum_{k} [ \overline{p_{k}\gamma_{\downarrow}}k(N-k+1)+\overline{p_{k}\gamma_{\uparrow}}(k+1)(N-k)] 
\eeq
where $\overline{p_{k}\gamma_{\downarrow}}=(\int^{\tau_{H}}_{0}dt p_{k}(t)\gamma_{\downarrow}^{H}+\int^{\tau_{C}+\tau_{H}}_{\tau_{H}}dt p_{k}(t) \gamma_{\downarrow}^{H})/(\tau_{H}+\tau_{C})$ denotes the short-hand notation of the time-average over one cycle. Similar relation holds for $\overline{p_{k}\gamma_{\uparrow}}$ as well. Here, we assume that the population of $k$-th symmetric Dicke state $|\psi_{k}^{\rm sym}\rangle$ follows an exponential form $\ln p_{k}\propto -k$. In this case, we find 
\beq
\bar{A}^{1\text{-ap}}=O(N). \label{sup_A1_scaling}
\eeq
Similarly, the analytical expression of $\bar{A}$ using $L^{3\text{-ap}}_{\omega_{0}}$ reads 
\beq
\bar{A}^{3\text{-ap}}=\omega_{0}^{2}\sum_{k} \left( \overline{p_{k}\gamma_{\downarrow}}k(N-k+1)[1+(k-1)(N-k)/2]^{2}+\overline{p_{k}\gamma_{\uparrow}}(k+1)(N-k)[1+k(N-k-1)/2]^{2}\right). 
\eeq
By assuming the form $\ln p_{k}\propto -k$, we find 
\beq
\bar{A}^{3\text{-ap}}=O(N^{3}). \label{sup_A3_scaling}
\eeq

Let us now recall the power-efficiency trade-off relation~\cite{Shiraishi19, Tajima21} [Eq.~(11) in the main text]:
\beq
\frac{P}{\eta_{\rm Car}-\eta}\leq c\bar{A}, \label{sup_PEtradeoff}
\eeq
where $\eta_{\rm Car}=1-\beta_{H}/\beta_{C}$ is the Carnot efficiency; $\beta_{H}$ and $\beta_{C}$ are the inverse temperatures of the hot and cold baths; $c=\beta_{C}\eta_{\rm Car}/[2(2-\eta_{\rm Car})^{2}]$ is a constant; and $\bar{A}=\tau^{-1}\int^{\tau}_{0}dt\sum_{\omega,k}\omega^{2} p_{k}c_{k}\mathcal{A}_{\omega}(\rho_{k},\{L_{a,\omega}\})$, where $\tau$ is the duration of time to complete one engine cycle. 
In Ref.~\cite{Tajima21}, the authors discussed that if $\bar{A}$ scales as $O(\mathcal{N}^{2})$, where $\mathcal{N}$ is the number of degeneracy, then the scaling $P=O(\mathcal{N})$ and $\eta_{\rm Car}-\eta = O(1/\mathcal{N})$ becomes possible. Moreover, an explicit model that realizes such scaling behavior has been introduced (see also Sec.~\ref{sup_sec_toric}). 
Similarly, if $\bar{A}$ scales as $O(N^{\alpha+1})$ and the efficiency scales as $\eta_{\rm Car}-\eta = O(1/N)$ as in Eq.~(\ref{sup_Carnot}), the power scales as $P=O(N^{\alpha})$ when the inequality in Eq.~(\ref{sup_PEtradeoff}) is saturated. Therefore, realizing a model with $\alpha \geq 1$ is practically important, because such scaling allows obtaining the power which is nonvanishing in the macroscopic limit ($N\rightarrow \infty$) while approximately attaining the Carnot efficiency~\cite{Tajima21}. Indeed, Fig.~\ref{sup_fig2} shows that for large $N$, the power scales as 
\beq
P^{3-\text{ap}}=O(N^{2}) \label{sup_Power_scaling}
\eeq
while the efficiency scales as \beq
\eta_{\rm Car}-\eta = O(1/N). \label{sup_efficiency_scaling}
\eeq

\revision{To realize this $O(N^{2})$ enhancement via $L^{3\text{-ap}}_{\pm \omega_{0}}$, the system-bath interaction Hamiltonian requires the form $H_{\rm int}=S\otimes B$, where $B$ is the bath operator and $S=\sum_{i}\sigma_{i}^{x}+\sum_{i< j< k}\sigma_{i}^{x}\otimes \sigma_{j}^{x}\otimes\sigma_{k}^{x}$ is the system operator. An interesting future direction is to find a setup that realizes this non-linear system-bath coupling, for example, using the nonlinear interactions~\cite{eriksson2024universal}, or the digital simulations of open quantum systems via trapped ion platforms~\cite{Muller_2011, RevModPhys.93.025001, PhysRevLett.129.063603, PRXQuantum.4.030311, PhysRevLett.129.063603, PRXQuantum.4.030311}.}

\section{\label{sup_sec_toric}Details on the permutation and even number bit flip invariance example}
In this section, we give further details on the permutation and even number bit flip invariant system discussed in the main text. 
The Hamiltonian reads
\beq
H = \epsilon \prod_{i=1}^{N}\sigma_{i}^{z}. \label{stabilizer}
\eeq
%When $N=4$, the above Hamiltonian describes a four-body interaction realizing a stabilizer operator that appears in Kitaev's toric code model. 
The Hamiltonian~(\ref{stabilizer}) has two eigen-energies $\pm \epsilon$, and the number of degeneracies are $\mathcal{N}_{\pm \epsilon}=\mathcal{N}=2^{N-1}$. The projection operators to the energy eigenspace read $
\Pi_{\pm\epsilon}=\frac{1}{2}(1\pm \prod_{i}\sigma_{i}^{z})$. 
The Hamiltonian~(\ref{stabilizer}) is permutation invariant, and it is also invariant under an even number bit flip operation
\beq
\{U_{g}\}=\{1,\sigma_{i_{1}}^{x}\sigma_{i_{2}}^{x},\sigma_{i_{1}}^{x}\sigma_{i_{2}}^{x}\sigma_{i_{3}}^{x}\sigma_{i_{4}}^{x}, \cdots\}. \label{bitflip}
\eeq
An explicit calculation reads $[H,U_{g}]=[\Pi_{\pm\epsilon},U_{g}]=0$.

\revision{
\subsection{\label{sec:Perm}Permutation symmetry and block diagonal structure}
In this subsection, we first consider only the permutation group $S_{N}$ and consider decomposing the energy eigenspace $\mathcal{S}_{\pm\epsilon}$. Using the irreducible representations of $S_{N}$, we have the block diagonal structure
\beq
\mathcal{S}_{k}=\bigoplus_{j=j_{\rm min}}^{j_{\rm sym}} \mathcal{H}_{j}^{k}\otimes \mathcal{K}_{j}^{k},
\eeq
where $j$ labels the irreducible representations of $S_{N}$, $k=\pm \epsilon$, $j_{\rm sym}=N/2$, and $j_{\rm min}=1/2$ if $N$ is odd and $j_{\rm min}=0$ if $N$ is even. The projection operator $\Pi_{\pm\epsilon}$ can be decomposed as
\beq
\Pi_{+\epsilon} = \bigoplus_{j=j_{\rm min}}^{j_{\rm sym}} \sum_{m}|j,m\rangle \langle j,m| \otimes I_{\mathcal{K}_{j}^{+\epsilon}}. \label{sup_Piplus1}
\eeq
If $N$ is even (odd), the summation over $m$ is taken only for even (odd) numbers within the range $m=j-N/2,\cdots,j+N/2$. We note that $\Pi_{-\epsilon}$ can be decomposed into a similar form. What should be noted here is that 
\beq
\dim(\mathcal{H}_{j}^{k}) \neq 1,
\eeq
and for example, $\dim(\mathcal{H}_{j_{\rm sym}}^{+\epsilon})=\lceil (N+1)/2 \rceil$. However, the subspace $\mathcal{H}_{j}^{k}$ is not invariant under the even number bit-flip operation, and thus it can be further decomposed. 

\subsection{Local and symmetric states}
The local energy eigenbasis for $\mathcal{S}_{\pm\epsilon}$ are given by ($V'_{g}=V_{g_{1}}U_{g_{2}}$ with $g=(g_{1},g_{2})$)
    \beqa
    \mathcal{B}_{+\epsilon}^{\rm loc}&=&\{ |\psi^{\rm loc}_{+\epsilon}(\alpha)\rangle \} = \{ V_{g}' |e\rangle^{\otimes N} \}_{g\in G}, \nonumber \\
    \mathcal{B}_{-\epsilon}^{\rm loc}&=&\{ |\psi^{\rm loc}_{-\epsilon}(\alpha)\rangle \} = \{ V_{g}' |g\rangle\otimes|e\rangle^{\otimes N-1} \}_{g\in G}.   \label{sup_stab_loc}
    \eeqa
In particular, we find that  
\beq
\frac{1}{|G|}\sum_{g}V_{g}|\psi^{\rm loc}_{\pm\epsilon}(\alpha)\rangle\langle \psi^{\rm loc}_{\pm\epsilon}(\alpha)|V_{g}^{\dagger} = \frac{1}{\mathcal{N}}\Pi_{\pm\epsilon}, \label{sup_stab_locdensity}
\eeq
and there is no nontrivial subspace in $\mathcal{S}_{\pm \epsilon}$ under the action of $V_{g}'$, meaning that $\dim(\mathcal{H}_{j}^{k})=1$ by considering $V_{g}'$. 

In addition, the symmetric state is uniquely given by $\rho_{\pm\epsilon}^{\rm sym}=|\psi^{\rm sym}_{\pm\epsilon}\rangle\langle \psi^{\rm sym}_{\pm \epsilon}|$, with
    \beq
    |\psi^{\rm sym}_{\pm\epsilon}\rangle = \frac{1}{\sqrt{2}} \left( |+\rangle^{\otimes N} \pm |-\rangle^{\otimes N} \right) ,
    \eeq
where $|\pm\rangle = (|e\rangle \pm |g\rangle)/\sqrt{2}$. 
One can check that it is invariant under the permutation $\{V_{g}\}_{g\in S_{N}}$ and Eq.~(\ref{bitflip}).  
By taking the superposition of local states~(\ref{sup_stab_loc}), we obtain that 
\beq 
|\psi^{\rm sym}_{\pm\epsilon}\rangle = \frac{1}{\sqrt{\mathcal{N}}}\sum_{g\in G}V_{g}' |\psi^{\rm loc}_{\pm\epsilon}\rangle .
\eeq

}

\subsection{Local and symmetric jump operators}

The symmetric jump operator is given by
    \beqa
    L^{\rm sym}_{2\epsilon} &=&  \sum_{m=0}^{\lceil N/2\rceil-1}\sum_{i_{1}<\cdots<i_{2m+1}} \Pi_{-\epsilon}\sigma^{x}_{i_{1}}\cdots \sigma^{x}_{i_{2m+1}}\Pi_{\epsilon} \nonumber \\
    &=&\Pi_{-\epsilon} \prod_{i}(\sigma_{i}^{x}+1) \Pi_{\epsilon} \nonumber \\
    &=& \mathcal{N} |\psi^{\rm sym}_{-\epsilon}\rangle\langle \psi^{\rm sym}_{\epsilon}|, 
    \eeqa
with 
\beq
    \mathcal{A}_{2\epsilon}(\rho^{\rm sym}_{\epsilon},L^{\rm sym}_{2\epsilon})= \mathcal{N}c_{\epsilon}^{\rm sym}  \label{sup_stab_scaling}
    \eeq
and $c_{\epsilon}^{\rm sym}=\gamma_{\downarrow}\mathcal{N}$. 
    
We can similarly take the first $2n+1$-body term approximation and find
    \beq
    L^{2n+1-{\rm ap}}_{2\epsilon}=\sum_{m=0}^{n}\sum_{i_{1}<\cdots<i_{2m+1}} \Pi_{-\epsilon}\sigma^{x}_{i_{1}}\cdots \sigma^{x}_{i_{2m+1}}\Pi_{\epsilon},
    \eeq
with 
\beq
    A_{2\epsilon}(\rho^{\rm sym}_{\epsilon},L^{2n+1-{\rm ap}}_{2\epsilon})=  \sum_{m=0}^{n} {}_{N}C_{2m+1} c_{k}^{2n+1-{\rm ap}}
    \eeq
and $c_{k}^{2n+1-{\rm ap}}=\gamma_{\downarrow}\sum_{m=0}^{n} {}_{N}C_{2m+1}$. 

Now, local jump operators can be obtained 
    \beq
    \{L^{2n+1\text{-loc}}_{a,2\epsilon}\}= \Bigl\{ \{ \Pi_{-\epsilon}\sigma_{i_{1}}^{x} \Pi_{\epsilon} \},\{ \Pi_{-\epsilon}\sigma_{i_{1}}^{x}\sigma_{i_{2}}^{x}\sigma_{i_{3}}^{x} \Pi_{\epsilon} \},\cdots, \{ \Pi_{-\epsilon}\sigma^{x}_{i_{1}}\cdots \sigma^{x}_{i_{2m+1}}\Pi_{\epsilon}\}  \Bigr\}.
    \eeq
These jump operators satisfy $
    A_{2\epsilon}( \rho^{\rm sym}_{\epsilon},L^{2n+1\text{-loc}}_{a,2\epsilon} )=c_{k}^{2n+1\text{-loc}}$ and $c_{k}^{2n+1\text{-loc}}=c_{k}^{2n+1-{\rm ap}} $.

Note that this permutation and even number bit flip invariant model corresponds to the $2\mathcal{N}$-state model discussed in Ref.~\cite{Tajima21} by noting that $A=O(\mathcal{N}^{2})$. Therefore, by applying the results discussed in Ref.~\cite{Tajima21}, we can realize a heat engine model that achieves $\eta_{\rm Car}-\eta=O(1/\mathcal{N})$ and $P=O(\mathcal{N})$. 

In Ref.~\cite{Kamimura23}, the authors impose a condition on the size of the system coupling operator $||S_{a}||$ and derive a bound on the heat current. On the other hand, our enhancement analysis is based on comparing different states and jump operators by excluding the difference of the normalization factor $c_{k}$. Note that we can renromalize the coefficient of the jump operators and set $c_{k}$ to be the same value in the above example as well.

\end{document}